\documentclass[twocolumn,trackchanges]{aastex7}
\usepackage{lmodern}
\usepackage{ae,aecompl,multirow}
\usepackage{threeparttable}
\usepackage{booktabs}
\usepackage{url}
\usepackage{amsmath}
\usepackage{graphics,comment}
\usepackage{textcomp} 
\usepackage{hyperref} 

\usepackage{booktabs}  
\DeclareUnicodeCharacter{2212}{\textminus}

\begin{document}

\title{Probing dust torus radius--luminosity relation: An WISE view}

\correspondingauthor{Suvendu Rakshit}

\author[0009-0009-9683-7940]{Ashutosh Tomar}
\altaffiliation{Deceased. Ashutosh Tomar (1998–2025) led this work but passed away before its completion. This paper is submitted in his memory.}
\affiliation{Aryabhatta Research Institute of Observational Sciences, Nainital\textendash263001, Uttarakhand, India}
\affiliation{Indian Institute of Technology Roorkee, Roorkee, Uttarakhand-247667, India}
\email{suvenduat@gmail.com}

\author[0000-0002-8377-9667]{Suvendu Rakshit}
\affiliation{Aryabhatta Research Institute of Observational Sciences, Nainital\textendash263001, Uttarakhand, India}
\email{suvenduat@gmail.com}

\author[0000-0001-9957-6349]{Amit Kumar Mandal}
\affiliation{Center for Theoretical Physics of the Polish Academy of Sciences, Al. Lotnik$\Acute{o}$w 32/46, 02\textendash668 Warsaw, Poland}
\email{amitmandal.ctp@pan.pl}

\author[0000-0002-4684-3404]{Shivangi Pandey}
\affiliation{Aryabhatta Research Institute of Observational Sciences, Nainital\textendash263001, Uttarakhand, India}
\affiliation{Department of Applied Physics/Physics, Mahatma Jyotiba Phule Rohilkhand University, Bareilly\textendash243006, India}
\email{shivangipandey@aries.res.in}


\begin{abstract}
We present measurements of the dusty torus sizes of 51 active galactic nuclei (AGNs) with a redshift of $z<$ 0.8. Our analysis utilizes about 16 years of optical photometric data of 146 AGNs from various time-domain surveys, including ASAS-SN, CRTS, and ZTF, along with 14 years of infrared data in the $W$1 ($\sim$ 3.4 $\mu$m) and $W$2 ($\sim$ 4.6 $\mu$m) bands obtained from the Wide-Field Infrared Survey Explorer (WISE). The estimated dust torus size ranges from 1000 to 3000 days, using both the cross-correlation analysis and lightcurve modeling through `MICA'. The measured lag has been corrected by $(1+z)^{-0.37}$, to account for cosmological time dilation and the torus temperature-gradient scaling. We conduct a linear regression analysis for both the $W$1 and $W$2 bands to examine the radius--luminosity ($R$--$L_{BOL}$) relationship under two conditions: one where the slope is fixed at 0.5 and one where it is allowed to vary. For the fixed slope of 0.5, we find the ratio of R$_{\mathrm{BLR}}$: R$_{W1}$: R$_{W2}$ to be 1: 9: 12, indicating that the torus lies outside the BLR and that its size increases with wavelength. Furthermore, we determine the relationship between torus size and L$_{BOL}$, yielding best-fit slopes of $0.413\pm0.047$ for the $W$1 band and $0.397\pm0.058$ for the $W$2 band. Both slopes are shallower than predicted by the dust radiation equilibrium model. Furthermore, our findings indicate that the torus size systematically decreases as the Eddington ratio increases, a trend that can be explained by the self-shadowing effects of slim disks.
\end{abstract}



\section{Introduction}
The Unification Model of Active Galactic Nuclei (AGNs) proposes that the observed differences among AGNs arise primarily from orientation effects with respect to the observer’s line of sight \citep{antonucci1985spectropolarimetry, miller1990spectropolarimetry}. A key feature of this model is a geometrically and optically thick dusty torus that obscures the central regions when viewed edge-on, leading to the classification of AGNs into two types: Type 1, which shows both broad and narrow emission lines, and Type 2, which exhibits only narrow lines due to obscuration of the broad-line region (BLR). The central engine comprises a supermassive black hole, an accretion disk, the BLR, the dusty torus, and the narrow-line region (NLR). The accretion disk fuels the AGN’s luminosity, while the BLR lies close to it and is often hidden by the torus. The NLR, located farther out, remains visible from all viewing angles. Thus, the dusty torus plays a crucial role in governing the observed spectral properties of AGNs.

In the near-infrared (NIR), the torus emission is primarily dominated by thermal radiation from hot dust grains, which are heated by ultraviolet and optical radiation from the accretion disk. This heating mechanism explains the characteristic infrared bump observed in AGNs, as described by the model of \citealt{barvainis1987hot}. The dusty torus extends spatially from $\sim$ 0.05 to 1 pc in AGNs with bolometric luminosities ranging between $10^{44}$ and $10^{47}$ $\mathrm{erg \, s^{-1}}$ \citep{mandal2024revisiting}. Due to its compact size, direct imaging of the torus is challenging. However, advances in NIR and Mid-Infrared (MIR) interferometry have enabled the successful resolution of the dusty torus in a limited number of nearby AGNs \citep{kishimoto2009exploring,kishimoto2011innermost,kishimoto2011mapping, burtscher2013diversity, dexter2020resolved, amorim2023toward}. Nonetheless, spatial resolution is often insufficient for high-redshift AGNs to resolve the torus. Instead, temporal resolution techniques, such as reverberation mapping \citep[RM;][]{blandford1982reverberation, peterson1993reverberation}, provide an alternative approach. RM relies on the time delay ($\tau$) between flux variations in the ionizing UV/optical continuum from the accretion disk and the subsequent response in the BLR or torus emission. By measuring this time lag, one can estimate the distance to the torus, as it corresponds to the time required for the UV/optical radiation to reach the torus and be reprocessed into NIR emission.

\begin{figure*}
    \centering
    \includegraphics[scale=0.4]{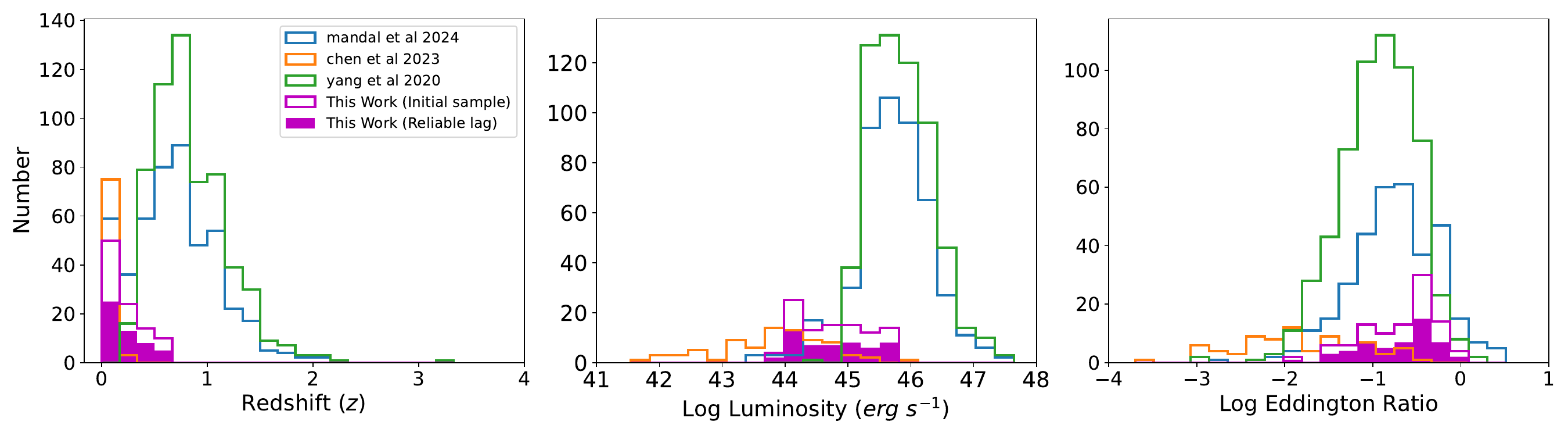}
    \caption{Comparison of the redshift, \textit{z} (left), bolometric luminosity (middle), and the Eddington ratio (right) distribution of the various samples with dust lag measurements along with our initial sample are shown in an empty magenta colored histogram, while the final sample of sources with reliable lag measurement in this work is shown in a filled histogram (see details in section \ref{ss:lag}).}
    \label{fig:comparison_plots}
\end{figure*}

Several studies have measured the inner extent of the dusty torus by investigating its correlation with luminosity \citep{suganuma2006reverberation, koshida2014reverberation, minezaki2019reverberation}. \citet{suganuma2006reverberation}, using a sample of four Seyfert 1 galaxies, found that $K$- band emission lagged behind $V$- band emission, consistent with the expected torus radius -- luminosity relation ($R_{\text{dust}}$--$L$), $R_{\text{dust}} \, \propto \, L^{\alpha}$ with $\alpha$ = 0.5, governed by the equilibrium between dust sublimation and irradiation from the accretion disk. Expanding the sample to 17 Seyfert 1 galaxies ($z < 0.05$), \citet{koshida2014reverberation} confirmed a similar correlation. However, \citet{minezaki2019reverberation} reported a shallower slope of $\alpha$ = 0.424, attributing it to anisotropic illumination from the accretion disk, which reduces the sublimation radius in the equatorial plane relative to polar regions \citep{kawaguchi2010orientation, kawaguchi2011near}. They also found that the torus contraction in response to a dimming accretion disk occurs over the years, leading to larger observed torus radii, particularly in lower-luminosity AGNs \citep{koshida2009variation}. Recently, \citet{2024MNRAS.531.3310L} determined the $K_S$ band dust lag measurement of 13 AGNs, finding systematically shorter lags than expected from the literature unless a wavelength-dependent correction factor is applied following \citet{minezaki2019reverberation}.

Focusing on the extent of the dusty torus, \citet{lyu2019mid} extended the wavelength coverage of dust RM studies from the $K$ band to the 3.4 $\mu$m ($W1$) and 4.6 $\mu$m ($W2$) bands of the Wide-field Infrared Survey Explorer (WISE). Using optical data from ground-based sky surveys, they analyzed a sample of 67 AGNs with redshifts $z < 0.5$ and found a correlation between torus radius and AGN luminosity, with slopes of 0.47 and 0.45 for the $W1$ and $W2$- bands, respectively. \citet{yang2020dust} also conducted a dust RM study on a significantly larger sample of 587 AGNs but did not analyze the correlation between torus radius and AGN luminosity. More recent investigations into this relationship include \citet{chen2023mid} and \citet{mandal2024revisiting}. \citet{chen2023mid} studied 78 AGNs with $z < 0.35$ using WISE data and found an even shallower slope in $R_{\text{dust}}$--$L$ of $\alpha = 0.36$ (0.37) for the $W1$ ($W2$) band. Additionally, they reported a negative correlation between the deviations from the best-fit $R_{\text{dust}}$--$L$ relation and the accretion rate \citep{du2014supermassive}, suggesting that an increase in accretion rate leads to a decrease in torus size due to the self-shadowing effect of the slim disk. \citet{mandal2024revisiting} further expanded the AGN sample significantly across a broader luminosity range ($10^{43.4} \, \text{erg \, s}^{-1} < L_{\text{BOL}} < 10^{47.6} \, \text{erg \, s}^{-1}$) and redshifts up to $z < 2$. They found that the $R_{\text{dust}}$--$L$ relation follows $R_{\text{dust}} \propto L_{\text{BOL}}^{0.39}$ for the $W1$ band and $R_{\text{dust}} \propto L_{\text{BOL}}^{0.33}$ for the $W2$ band. Furthermore, they identified a moderate negative correlation between deviations from the best-fit $R_{\text{dust}}$ – $L_{\text{BOL}}$ relation and the Eddington ratio, reinforcing the influence of accretion rate on torus structure.

Recent BLR-RM campaigns targeting super-Eddington sources \citep{2016ApJ...825..126D, 2018ApJ...856....6D} have revealed significantly shorter lags, by factors of 2 to 6 than those predicted by photoionization models \citep{1972ApJ...171..213D}, as well as deviate from the general trend of the BLR radius -- luminosity ($R_{\text{BLR}}$--$L$) relation  \citep{2024ApJS..275...13W}. This discrepancy suggests a more complex BLR geometry in these AGNs. In highly accreting systems, slim accretion disks generate strong self-shadowing effects, leading to anisotropic radiation fields and distinct BLR structures \citep{2014ApJ...797...65W}. Moreover, the inner dust torus in such AGNs may contract, potentially approaching the self-gravity radius of the disk \citep{2011MNRAS.417.2562K, 2014ApJ...793..108W}. Since super-Eddington sources already deviate from the $R_{\text{BLR}}$--$L$ relation, a similar deviation from the $R_{\text{dust}}$--$L$ relation is also expected. To improve our understanding of the dust torus radius -- luminosity relationship, we analyze a sample of over 350 AGNs with $z < 0.8$, including 81 strong Fe II emitters with Fe II to H$\beta$ equivalent width ratios $>$ 1, and measure reliable lag of about 50 AGNs. Note that the redshift limit in the initial sample is imposed to include only H$\beta$-based black hole masses. Additionally, depending on redshift, the W1 (W2) band emission includes dust contribution from the W1 to H-band (W2 to K-band). The structure of this paper is as follows: In Section \ref{ss:sample}, we describe our sample and data acquisition. Section \ref{ss:lag} details the time-series analysis and lag estimation. In Section \ref{ss:results}, we present results focusing on the correlations between AGN parameters and the dust lag, followed by a discussion. Finally, we summarize our main findings in Section \ref{ss:sum}.

\begin{table*}
    \centering
        \caption{An overview of the surveys employed in this work.}
         \label{surveytab}
    \begin{tabular}{cccccc}
    \hline
    \hline
        Survey  & Filter & Duration & Cadence(days) & Coverage & Depth(mag) \\
        (1) & (2) & (3) & (4) & (5) & (6) \\
        \hline
        CRTS & unfiltered & 2005--13 & $\sim$20  & 33,000 deg$^2$ & 19--21\\ 
        ASAS-SN & $V$, $g$ & 2012--24 & $\sim$3-4 & All sky & 17V, 17.5g\\
        ZTF & $r$ & 2018--2024 & $\sim$3-4 & 25,000-30,000 deg$^2$ & 20.6 \\
        WISE/NEOWISE & W1,W2 & 2010-2024 & $\sim$ 180 & All sky & 16.6(W1), 16.0(W2)\\
        \hline
    \end{tabular}
    \vspace{0.1cm}
   
    \raggedright  Note: Columns are as follows (1): name of the survey, (2): filter used, (3): duration of light curve observations, (4): typical cadence of the light curve, (5): sky coverage of the survey, and (6): limiting magnitude.
\end{table*}

\section{Sample and Data}
\label{ss:sample}

To investigate the $R_{\text{dust}}$–$L$ relation using the dust-RM (DRM) technique, we selected our parent sample from the catalog compiled by \citet[][hereafter R20]{rakshit2020spectral}, which is based on the 14th data release of the Sloan Digital Sky Survey (SDSS$-$DR14) quasar catalog. This catalog contains photometric and spectroscopic measurements for 526,265 AGNs. To refine our sample, we applied selection criteria of redshift $z < 0.8$, the signal-to-noise ratio of the spectral continuum  $>$ 10, and g-band magnitude $<$ 17, resulting in a sample of 352 objects. Among these, 81 AGNs are classified as strong Fe II emitters or highly accreting systems, defined by $\mathcal{R}_{Fe II} >1 $, where $\mathcal{R}_{Fe II}$ is the ratio of the equivalent widths of the Fe II complex (4435$-$4685 {\AA}) to the H$\beta$ line. Therefore, our initial parent sample comprises 352 AGNs, including 81 highly accreting AGNs with $\mathcal{R}_{Fe II} >1 $.

We obtained the spectral properties of our selected objects from the R20 catalog, which provides measurements of key parameters, such as the luminosity at 5100 {\AA} ($L_{5100}$), black hole mass, and the continuum and line properties of various emission lines, including H$\alpha$, H$\beta$, H$\gamma$, Mg II, L$\alpha$, and Fe II. The catalog covers a broad range of bolometric luminosities, spanning $44.4 < \log L_{\text{BOL}} < 47.3 \, erg \, s^{-1}$. To place our sample in context, Figure \ref{fig:comparison_plots} presents the redshift, luminosity, and Eddington ratio distributions compared to previous studies. While our sample includes comparatively lower-luminosity AGNs than those analyzed by \citet{yang2020dust} and \citet{mandal2024revisiting}, it focuses explicitly on highly accreting AGNs, providing a unique perspective on this population. Note that our final sample of sources with reliable lag measurements, shown by the filled magenta histogram, is much lower than our initial sample (empty magenta histogram) (see section \ref{ss:lag}).

\subsection{Optical Data}

To obtain light curve data for our sample, we first searched various transient surveys, beginning with the All-Sky Automated Survey for Supernovae (ASAS-SN)\footnote{\url{https://asas-sn.osu.edu/}}. ASAS-SN monitors the entire visible night sky using 24 telescopes worldwide, achieving a depth of 17 magnitudes in the $V-$ band with a cadence of 2$-$3 days \citep{shappee2014man, kochanek2017all}. It provides photometric data in the $V-$ band (limiting $\sim$17 mag, 2012$-$2018) and the $g-$ band (limiting $\sim$17.5 mag, late 2017$-$present). We queried the ASAS-SN Photometry Database for light curve data and found that 347 out of 352 objects, including 80 highly accreting AGNs, have available $V$ and $g-$ band light curves. The reported ASAS-SN magnitudes were converted into fluxes using the conversion factors provided by \citealt{bessell1998model} and \citealt{fukugita1996sloan}.

\begin{figure}
    \centering
    \includegraphics[width = 9cm, height =7cm]{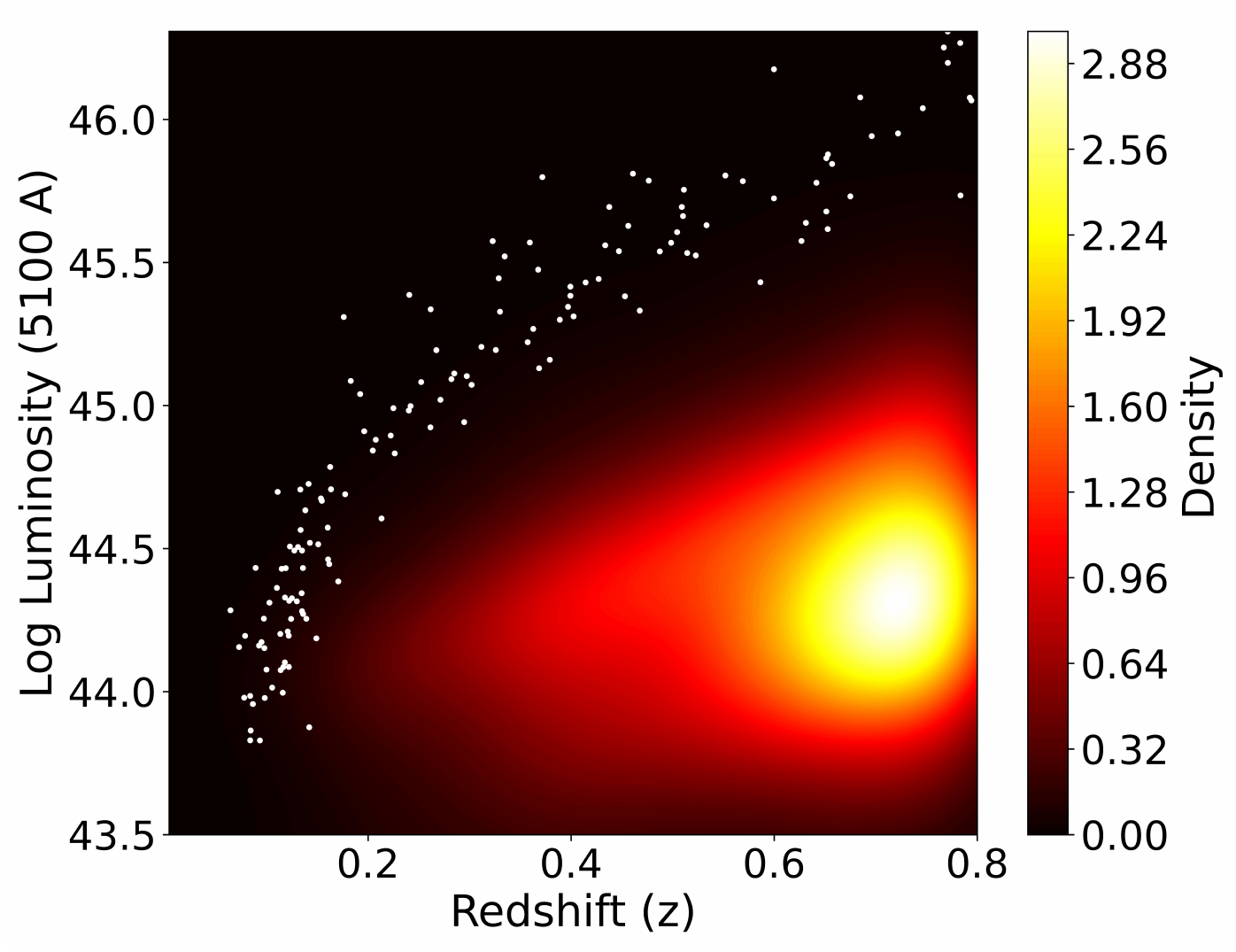}
    \caption{Luminosity--redshift density map of the total number of objects selected from \citealt{rakshit2020spectral} within redshift $z$ $<$ 0.8, SNR $>$ 10, and g-magnitude $<$ 17. The final sample of 146 objects selected for lag analysis is shown as white dots.}
    \label{fig:density_plot}
\end{figure}

Additionally, we obtained $V-$ band photometric light curve data from the Catalina Real-Time Transient Survey (CRTS)\footnote{\url{http://crts.caltech.edu/}}. Established in late 2007, CRTS was designed as a synoptic survey for optical transients, utilizing publicly available Catalina Sky Survey (CSS) image survey data. Covering 33,000 square degrees of the sky, it aimed to identify rare and interesting transient phenomena. The survey operated from 2005 to 2013, achieving a photometric depth of 19$-$21 $V-$ mag with an average cadence of approximately three weeks. Although the photometric data are unfiltered, they are broadly calibrated to the Johnson $V$ band using 2MASS data \citep{drake2009first}. With the exception of 7 targets, 340 targets (including 79 highly accreting AGNs) have CRTS data.

Further, we collected data from the Zwicky Transient Facility (ZTF)\footnote{\url{https://irsa.ipac.caltech.edu/}}, a wide-field sky survey that utilizes a new camera mounted on the Samuel Oschin Telescope at the Palomar Observatory in California. Commissioned in 2018, ZTF supersedes the (intermediate) Palomar Transient Factory (2009$-$2017), which used the same observatory. Designed to detect transient objects with rapid changes in brightness, ZTF observes in visible wavelengths to cover an area of 47 square degrees. ZTF provides data in the $g$, $r$, and $i-$ bands with median depths of $g \sim$ 20.8 mag,  $r \sim$ 20.6 mag,  $i \sim$ 19.9, with a typical observation cadence of 3 days. For our analysis, we selected ZTF data from data release (DR) 20 covering data between March 2018 to Oct 2023 with 'catflags' = 0 and converted magnitudes into flux using the zero-points provided by the SVO filter profile service\footnote{\url{http://svo2.cab.inta-csic.es/theory/fps/index.php?mode=browse&gname=Palomar&gname2=ZTF&asttype=}}. Since the $r-$ band contains the most data points, we used only $r-$ band data from ZTF. We note that for $z<0.8$, contamination from H$\beta$/H$\alpha$ line emission in the g and r bands is $\leq$ 20\% \citep{2016ApJ...821...56F}, and the relatively large PSF \cite[median FWHM $\sim2^{\prime\prime}$;][]{10.1093/mnras/stab3093} can include host starlight; however, by selecting AGNs with clear variability ($F_{var} >\sigma_{var}$; see section \ref{sec:cali}), these effects have negligible impact on the long-timescale dust lags analyzed in this work. The availability of ZTF data further refined our sample size, resulting in 330 objects with 77 high-accreting AGNs. The survey details are provided in Table \ref{surveytab}.

\subsection{IR data from WISE/NEOWISE}

The Wide-field Infrared Survey Explorer (WISE) was an infrared-wavelength astronomical space telescope active from December 2009 to February 2011. In September 2013, the spacecraft was reassigned to a new mission known as NEOWISE (Near Earth Object Wide-field Infrared Survey Explorer). During its original mission, WISE surveyed the full sky in four infrared wavelength bands: 3.4, 4.6, 12, and 22 $\mu$m, until its hydrogen cooling was depleted in September 2010. Following this, the NEOWISE program began mapping the sky from August 2010 to February 2011, focusing on detecting near-Earth objects. In December 2013, the NEOWISE Reactivation commenced, scanning the sky in the W1 and W2$-$ bands with a sampling interval of approximately six months. Each visit in the WISE/NEOWISE missions typically included 10$-$20 observations within a 36$-$hour window. For our AGN sample, we collected infrared photometric light curves from the WISE/NEOWISE missions, specifically in the W1 (3.4 $\mu$m) and W2$-$ (4.6 $\mu$m) bands. The effective rest-frame wavelengths of these bands range from 2.6--3.4 $\mu$m for W1 and 3.5--4.6 $\mu$m for W2 within our sample. We selected only high-quality data, applying stringent criteria including flags for the best image quality (qi\_fact = 1), larger South Atlantic Anomaly separations (saa\_sep $\ge$ 5), no contamination from scattered moonlight (moon\_masked = 0), and exclusion of spurious detections (cc\_flags = 0).

We collected IR data in W1 and W2 bands for all the 330 AGNs of our sample. To convert magnitudes into flux densities, we used the zero-point flux densities:
 f$_\lambda$(W1 = 0) = 8.03 $\times$ 10$^{-12}$ erg cm$^{-2}$ s$^{-1}$ and A$^{-1}$ and f$_\lambda$(W2 = 0) =  2.43 $\times$ 10$^{-12}$ erg cm$^{-2}$ s$^{-1}$ and A$^{-1}$. We then took the weighted average flux in those 36$-$hour windows by
    \begin{equation}
        f_{epoch} = \frac{1}{N} \; \frac{\sum_{i=1}^{N} \frac{f_i}{\sigma_i}}{\sum_{i=1}^{N} \frac{1}{\sigma_i}}
        \label{weifluxbin}
    \end{equation}
     and 
    \begin{equation}
        \sigma_{epoch} = \frac{1}{N(N-1)} \; \sum_{i=1}^{N} (f_i - \Bar{f}_{epoch})^2 + \frac{1}{N^2} \sum_{i=1}^{N} \sigma_i^2
        \label{weifluxerrbin}
    \end{equation}
     where N is the Number of data points within a single window, f$_i$ is the flux of the i$^th$ observation, $\sigma _i$ is the uncertainty of f$_i$, and 
     \begin{equation}
         \Bar{f}_{epoch} = \frac{1}{N} \sum_{i}^{N} f_i
     \end{equation}

\begin{figure*}
    \centering
       \includegraphics[width=1\linewidth, height=0.55\linewidth]{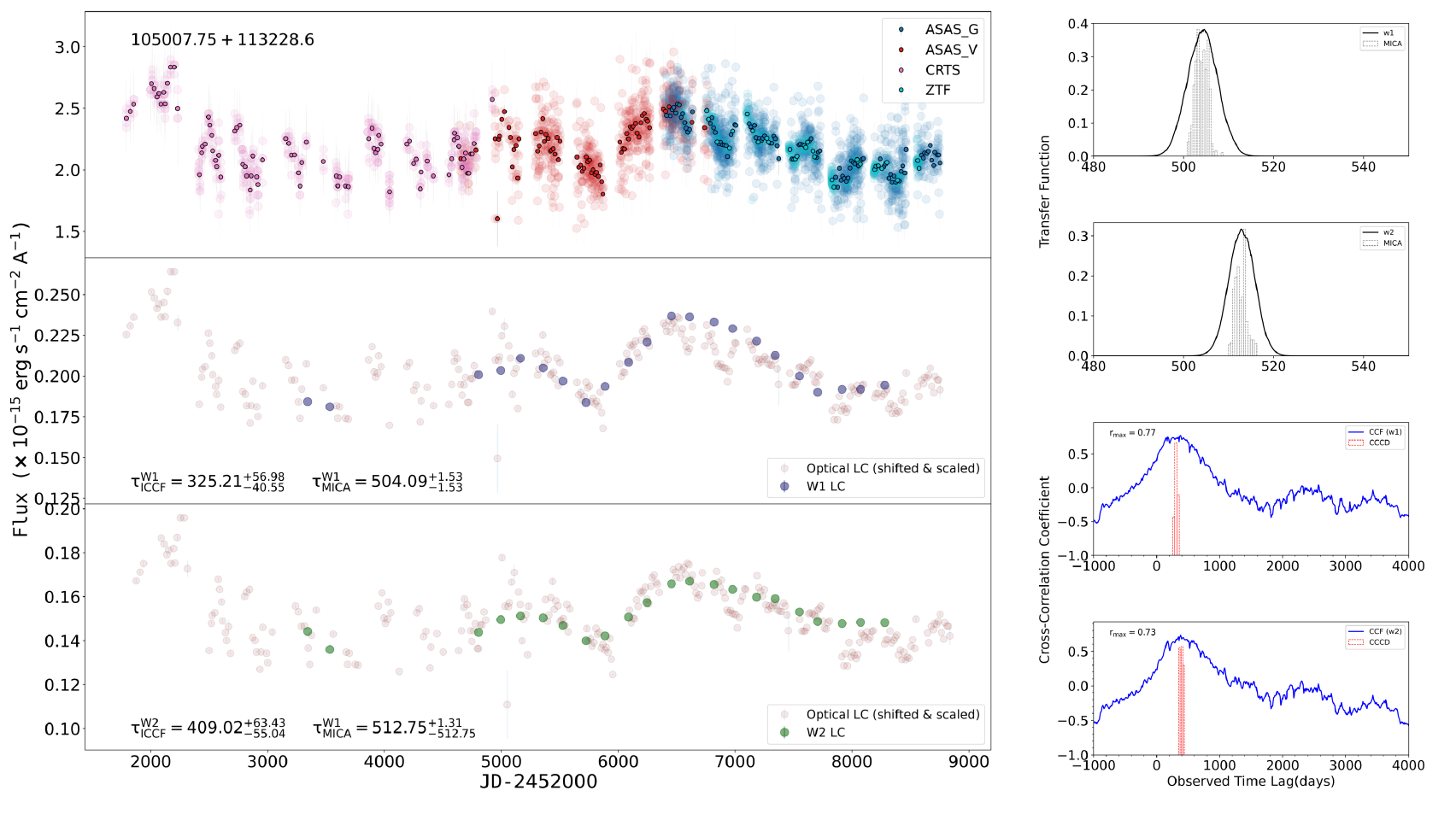}
       \includegraphics[width=1\linewidth, height=0.55\linewidth]{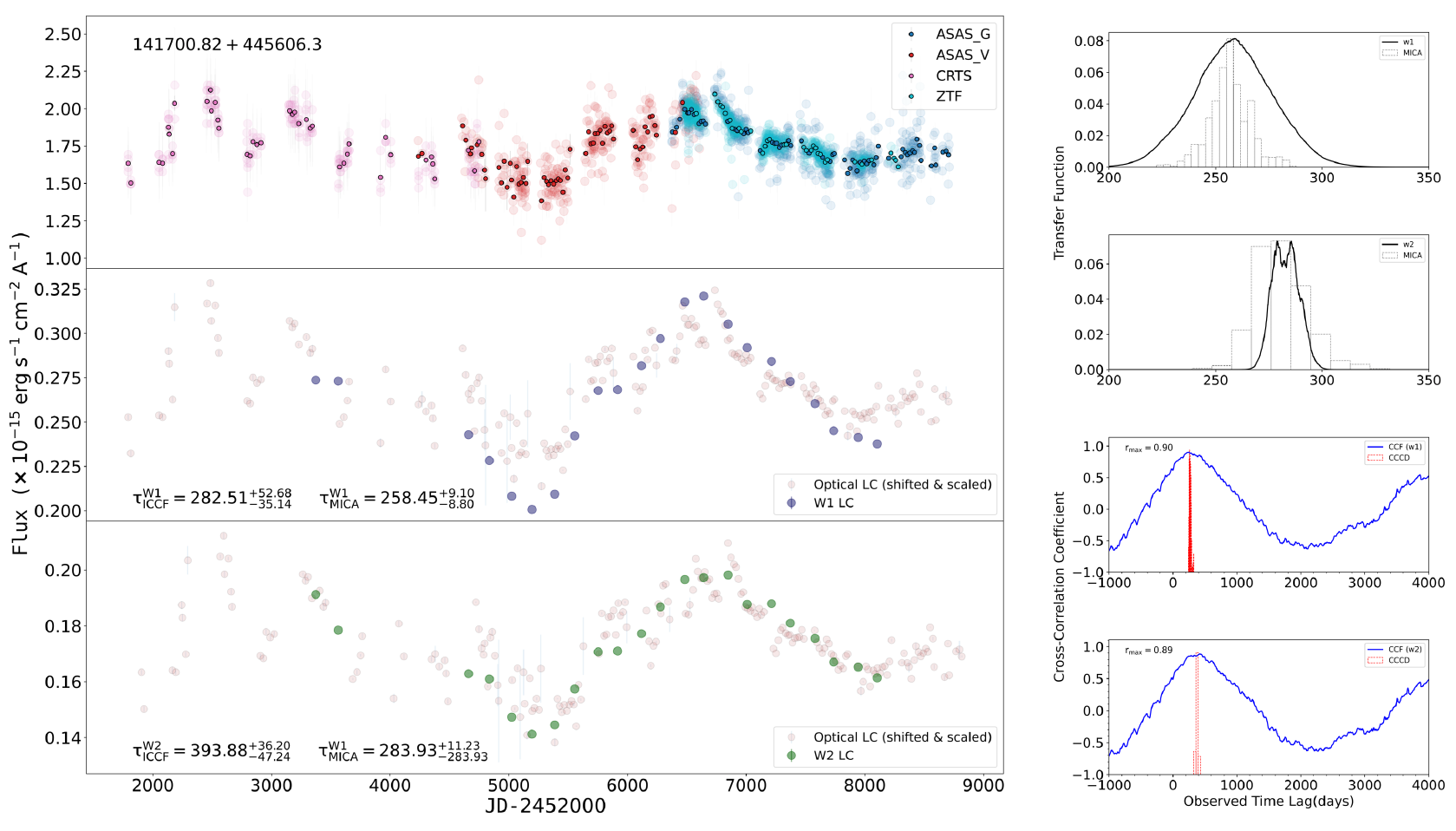}
    \caption{The light curves of two objects of our sample viz, 105007.75+113228.6 (top) and 141700.82+445606.3 (bottom), where the left panel shows the calibrated unbinned (with lower contrast) and binned optical light curves (upper) and IR light curves for the W1 (middle) and W2 (bottom) bands, respectively. In the lower two panels of the light curves, the optical light curve has been scaled and shifted by the lags found by ICCF. The Right panel illustrates the transfer function and the lag distribution found by employing {\tt MICA} and the CCF and CCCD found by ICCF.}
    \label{fig:lightcurves}
\end{figure*}

\subsection{Calibration of Optical Light Curves.}\label{sec:cali}

The optical light curve data used in this study are sourced from different surveys, each of which employs its own photometric data reduction pipeline. Since the data comes from various bands and surveys, it is essential to calibrate the data to account for systematic offsets between them. 

To resolve this, we employed the Bayesian package {\tt PyCALI}\footnote{\url{https://github.com/LiyrAstroph/PyCALI}} \citep{li2014bayesian}. {\tt PyCALI} is a Python-based tool that uses a damped random walk \citep[DRW;][]{kelly2009variations, macleod2010modeling} to model light curves and determine the optimal scaling factors for inter-calibration. Specifically, it calculates a multiplicative factor $\varphi$ and an additive factor \textit{G}, allowing for the calibration of the light curves using the relation \textit{F = $\varphi$ F$_{obs}$ + G}. This is achieved by exploring the posterior distribution with a diffusive nesting algorithm \citep{brewer2011diffusive}.

In our analysis, we chose the ASAS-SN $V-$ band data as the reference light curve and used it to calibrate the light curve data from other surveys for each object in the sample. Following this, we utilized the flux density measurements and their corresponding errors (which account for measurement errors and systematic errors), as provided by {\tt PyCALI}, for further analysis. To ensure consistent sampling across the light curves, the calibrated data were binned with a bin-width of 15 days, using equations \ref{weifluxbin} and \ref{weifluxerrbin}. Additionally, we applied three-sigma clipping to remove any outlier points, ensuring the quality and consistency of the final light curves. Inter-band continuum delays \citep[$<$10 days;][]{2017ApJ...836..186J} were neglected during inter-calibration, as they have negligible impact on the much longer IR delays.

Next, we calculated the mean fractional variations (F$_{var}$) and the flux variance ($\sigma_{var}$) to assess the quality of the binned and inter-calibrated optical light curves \citep{peterson2001variability}. To ensure that only reliable objects with considerable flux variability were included in the further analysis, we applied the condition: F$_{var}$ $>$ $\sigma_{var}$. After this selection, we were left with a sample of 146 objects, including 30 high-accreting AGNs. The density plot of the parent sample objects in the catalog and the final sample of 146 objects selected for lag analysis is shown in Fig. \ref{fig:density_plot}. In the left panels of Fig. \ref{fig:lightcurves}, we present the final inter-calibrated optical light curves, along with the IR light curves in the $W$1 and $W$2$-$ bands, for two targets from our sample.

\section{Time Series Analysis}
\label{ss:lag}

We analyzed optical light curves of all 146 objects from 2008 to 2024 with a cadence of 15 days, alongside IR data from WISE and NEOWISE in the $W$1 and $W$2$-$ bands, which have a 180-day cadence spanning between 2010 to 2024, to determine the lags between the optical and different IR light curves.

\subsection{Lag Estimation}

Since the IR emission from the torus echoes the optical emission from the accretion disk, both emissions are synchronous, with a delay/lag observed in the IR emission. To determine these lags, we employed two different methods: the Interpolated Cross-Correlation Function \citep[ICCF;][]{gaskell1986line,gaskell1987accuracy}, and Multiple and Inhomogeneous Component Analysis\footnote{\url{https://github.com/LiyrAstroph/MICA2}} \citep[{\tt MICA};][]{li2016non}. For ICCF, we used the Python code {\tt PyCCF} \citep{2018ascl.soft05032S}, which calculates the cross-correlation coefficient at each time step in the given lag range by linearly interpolating the light curves. For this analysis, we chose a lag range corresponding to approximately 70$\%$ of the baseline defined by the optical light curve, which is sufficient for lag measurements \citep{grier2017sloan, grier2019sloan, yang2020dust, mandal2024revisiting}. The lag was then determined by identifying the peak of the correlation coefficient distribution within this range and calculating the centroid around the most significant peak, using a cutoff of 80$\%$ of the maximum CCF value (r$_{max}$) as defined by \citealt{peterson1998uncertainties}:

\begin{equation}
    \centering
        \tau_{cent} = \frac{\Sigma_j\;\tau_j\;CCF_j}{\Sigma_j\;CCF_j}
\end{equation}

The uncertainties in the lags, on the other hand, were calculated using Monte Carlo simulations based on flux randomization (FR) and random subset selection (RSS) \citep{peterson2004central, peterson1998uncertainties} methods. These uncertainties were estimated using the 15.87th and 84.13th percentiles of the posterior distribution of the cross-correlation centroid, which correspond to the 1$\sigma$ error in a Gaussian distribution.

\begin{figure*}
    \centering
    \includegraphics[width=1.\linewidth]{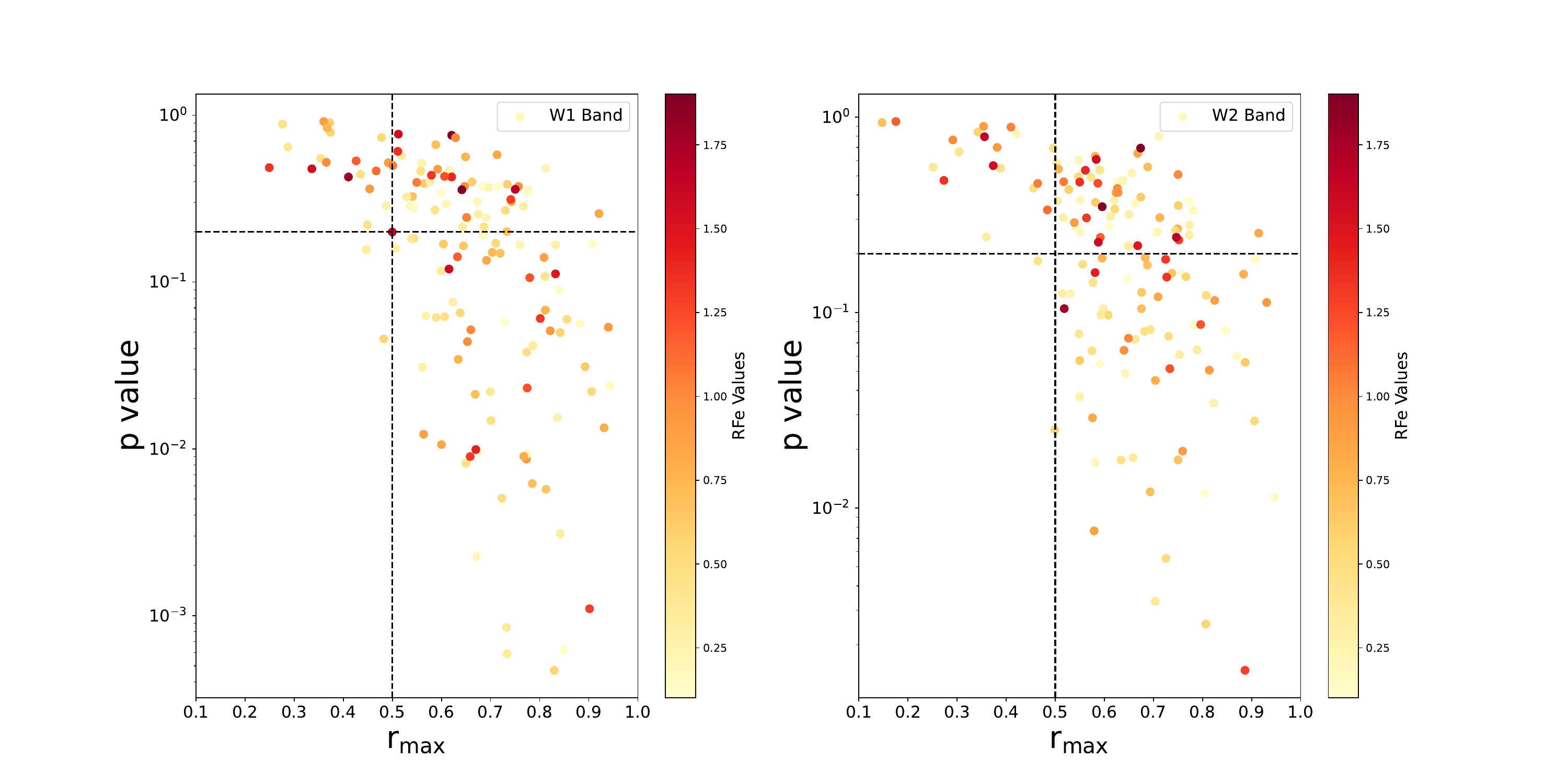}
    \caption{Quality assessment through p-valve and maximum cross-correlation coefficient (r$_{max}$) color coded by $\mathcal{R}_{Fe II}$ values. The selection criteria are p-valve $<$ 0.2 and r$_{max}$ $>$ 0.5, as shown by the horizontal and vertical dashed lines, respectively.}
    \label{fig:py2ccf}
\end{figure*}

We also used {\tt MICA} to estimate the lags between the optical and IR light curves. The {\tt MICA} models the light curves using DRW, characterized by two parameters: amplitude ($\sigma_d$) and the time-scale of variability ($\tau_d$). It then retrieves the transfer function as a family of displaced Gaussians. To estimate the lag uncertainties, we applied the Markov Chain Monte Carlo approach, which finds the median of the transfer functions and uses the 68.3$\%$ quantile to determine the uncertainty. The same lag range was applied to the objects as used in the ICCF method.

\begin{figure}
    \centering
    \includegraphics[width=9cm, height=9cm]{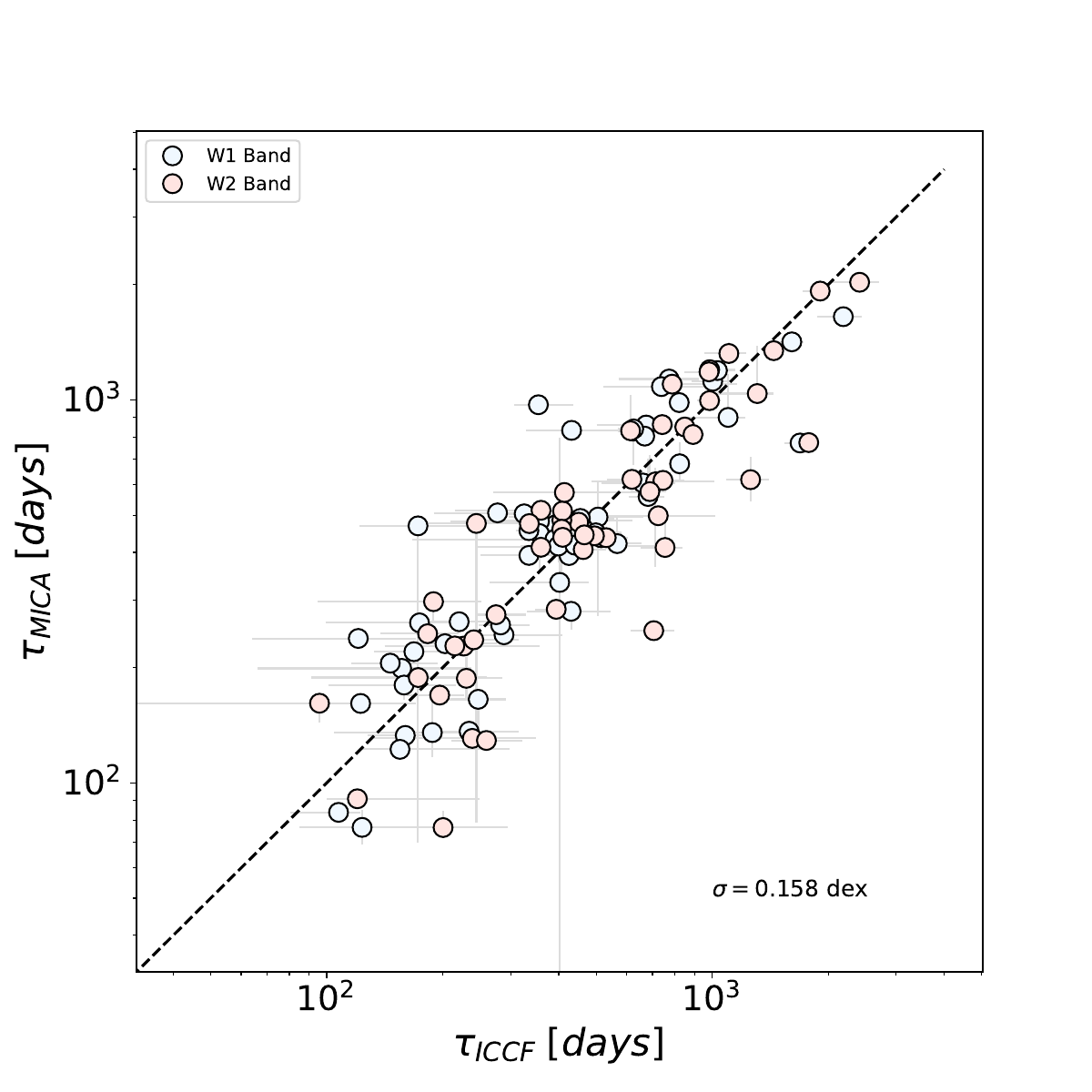}
    \caption{Comparison of the lags estimated by ICCF and MICA. The dashed black line shows 1:1 relation. The measured lags show good agreement with the 1:1 relation with an intrinsic scatter of $\sim$ 0.158 dex. }
    \label{fig:lag_comparison}
\end{figure}

\begin{figure}
    \centering
\includegraphics[width=8.5cm,height=8cm]{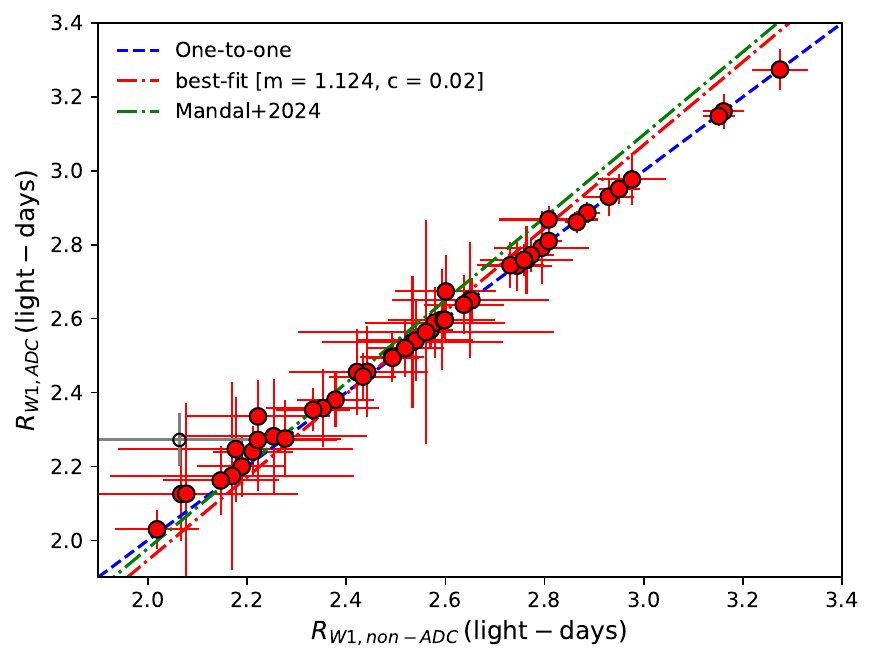}
\includegraphics[width=8.5cm,height=8cm]{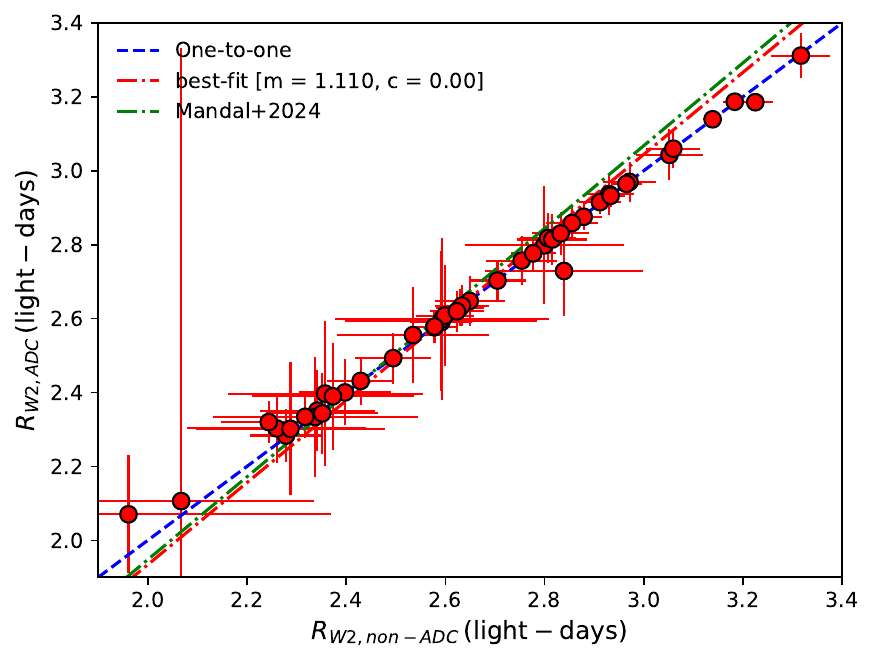}
    \caption{The comparison of dust lag after correction for accretion disk contribution (y-axis) vs. before (x-axis). The one-to-one slope is marked by the dashed blue line along with our best-fit shown by the dashed red line, which closely matches that obtained by \citet{mandal2024revisiting} represented by the dashed green line.}
    \label{fig:adc}
\end{figure}

\subsection{Quality Assessment of the Lags.}

In this section, we assess the quality of the lag measurements obtained from ICCF analysis. Given that WISE/NEOWISE data are sparsely sampled and the optical light curves are derived from different surveys with independent calibration methods, various sources of uncertainty may affect the resulting lags. A key parameter used to evaluate the correlation strength between light curves in CCF analysis is the maximum cross-correlation coefficient, r$_{max}$. However, r$_{max}$ can peak at a high value when specific features in the two light curves align, even if only a few data points are available. Consequently, relying solely on r$_{max}$ may be insufficient to assess the reliability of the measured lags. 

To enhance the robustness of our analysis, we employed the publicly available Python package PyI$^{2}$CCF\footnote{\url{https://github.com/legolason/PyIICCF}}, which incorporates an additional reliability check for detected lags. PyI$^{2}$CCF implements the ICCF method while testing the null hypothesis to determine the probability of obtaining an r$_{max, sim}$ value computed from two uncorrelated, simulated light curves that exceeds the observed r$_{max, obs}$. The simulated light curves are generated with the same cadence and signal-to-noise ratio (S/N) as the observed data. PyI$^{2}$CCF offers two interpolation methods: linear and non-linear. Since ICCF inherently relies on linear interpolation, we adopted the same approach within PyI$^{2}$CCF to ensure consistency. For each object, we generated mock optical and infrared light curves based on the best-fit damped random walk (DRW) models. A synthetic light curve, 100 times longer than the observed baseline, was first produced, from which random segments matching the observed duration were extracted. Gaussian noise, scaled to the observational uncertainties, was then added, and the resulting curves were downsampled to reproduce the cadence and temporal coverage of the data. We generated 2000 mock lightcurves per object and computed the probability p(r $>$ r$_{max}$) as the fraction of simulated cases where r$_{max}$ exceeded the observed r$_{max}$. Figure \ref{fig:py2ccf}  shows the p-value as a function of r$_{max}$ for all the targets. Finally, we applied selection criteria of p $<$ 0.2 and r$_{max}$ > 0.5 to filter out unreliable lag measurements.  With the above criteria, we identified 69 objects in $W$1$-$band and 64 in $W$2$-$band. To refine our sample further, we applied the following criteria to select targets with reliable lag: (i) the objects with positive lags, (ii) the objects having lag values lower than 3500 days, which is about $70\%$ of IR light curve baseline, and (iii) the objects having $ 0.33 < \tau_{mica}/\tau_{ICCF} < 3$.

As a result, out of the  146 objects, the final selection comprised 51 objects for the W1$-$ band and 47 objects for the W2$-$ band. The redshift, luminosity, and Eddington ratio distributions of the final sample of 51 objects with reliable lag measurement in the W1 band are shown in Figure \ref{fig:comparison_plots}. Figure \ref{fig:lag_comparison} presents a comparison of the lag estimates obtained using ICCF and {\tt MICA}, revealing a scatter of 0.158 dex around the 1:1 relation. Finally, we visually inspected the lags obtained by the ICCF by shifting the optical light curve, which is in agreement with the IR light curves.

\subsection{Correction for the redshift in the lags measured.}
For the high redshift AGNs, it becomes necessary to consider two redshift effects viz, the cosmological time dilation, where a factor of (1+\textit{z})$^{-1}$ comes, and another is the correction for effect due to temperature gradient of the dusty torus which is anisotropic in the IR wavelength range (\citealt{oknyanskij2001reverberation},\citet{yoshii2014new},\citealt{minezaki2019reverberation},\citealt{figaredo2020dust},\citealt{chen2023mid},\citealt{mandal2024revisiting}).

Adopting the methodologies from \citealt{yoshii2014new} and \citealt{minezaki2019reverberation}, we determine the wavelength-dependent correction factor as (1 +\textit{z})$^\gamma$,  where $\gamma$ is defined as $\gamma = \rm log(\tau_{W2}/\tau_{W1})/ \rm log(\lambda_{W2}/\lambda_{W1})$. From our final sample of 51 objects in the W1 band and 47 objects in the W2 band, we identified 45 common objects across both bands. For these objects, we calculated the lag ratio $\tau_{\mathrm{ratio}} = \tau_{W2}/\tau_{W1}$ and subsequently computed $\gamma$. Our results show a median lag ratio of $\tau_{\mathrm{ratio,median}}$=1.21$\pm$0.06 for the ICCF lags, which is in agreement with the findings of \citealt{lyu2019mid} ($\tau_{\mathrm{ratio,median}}$ = 1.15), \citealt{chen2023mid} ($\tau_{\mathrm{ratio,median}}$ = 1.26) and \citealt{mandal2024revisiting} ($\tau_{\mathrm{ratio,median}}$ = 1.20). Utilizing our obtained $\tau_{\mathrm{ratio,median}}$, we estimated $\gamma$ $\sim$ 0.63, which is consistent with the value found by \citealt{mandal2024revisiting} ($\gamma \sim \rm 0.62$). Additionally, applying the cosmological time dilation correction factor, the combined correction factor for rest-frame observations becomes (1+\textit{z})$^{-1+\gamma}$ = (1+\textit{z})$^{-0.37}$. This correction factor is then applied to the final selected sample, and the corrected time lags are used in subsequent analyses.

\begin{table*}
\centering
\movetableright= -50mm
\caption{Details of the object's properties and lag measurement results obtained by the ICCF method. A portion of the table is shown for guidance; the complete table is available online in a machine-readable format.}
\label{tab:lag-result}
\resizebox{18cm}{!}{
\fontsize{16pt}{16pt}\selectfont
    \begin{tabular}{llllllllllr}
        \hline
        \hline
        	Name &	RA &	DEC	& $z$ &	$\log L5100$	& $\log M_{BH}$	& $\log R_{EDD}$ & $\tau$ (W1)  &	$r_{max}$ (W1) &	$\tau$ (W2) &	$r_{max}$ (W2) \\
                & (deg) & (deg) &     & (erg s$^{-1}$) & ($M_{\odot}$) &  & (days) &  & (days) & \\ 
            (1) & (2) & (3) & (4) & (5) & (6) & (7) & (8) & (9) & (10) & (11) \\\hline          
015950.24+002340.8 &	29.959341 & 0.394696 &	0.1627	& 44.785 &	8.125 &	-0.51 & $392^{+44}_{-39}$	& 0.83 &	$450^{+54}_{-51}$ &	0.81 \\
024240.31+005727.1 &	40.667992 &	0.957531 &	0.5690 &	45.784 &	8.898 &	-0.287 &	$1005^{+116}_{-119}$ &	0.80 &	$1102^{+124}_{-154}$	& 0.78 \\
005205.56+003538.1 &	13.023225 &	0.593944 &	0.3993 &	45.415	& 8.426 & -0.184	& $507^{+157}_{-210}$	& 0.60 & $711^{+300}_{-223}$	& 0.55 \\
        \hline
    \end{tabular}
    }
\parbox{\linewidth}{
        \noindent Note: Columns are as follows. (1): Name of source, (2): RA, (3): Dec., (4): redshift, (5): luminosity at 5100 \AA\, ($L_{bol}=9.26\times L_{5100}$), (6): logarithmic black hole mass, (7) logarithmic Eddington ratio, (8) observed frame time lag in W1, (9) maximum correlation coefficient in W1, (10) observed frame time lag in W2, and (11) maximum correlation coefficient in W2. The values in columns (4) to (6) are taken from \citet{rakshit2020spectral}.}
\end{table*}

\subsection{Correction for the accretion disk contamination}
The accretion disk contributes to the IR flux, making the lag measurement shorter than the true lag. Therefore, it is essential to remove any contamination from the accretion disk. \citet{koshida2014reverberation} suggested a formal to estimate the accretion disk contribution in the measured IR flux as
\begin{equation}
    \mathrm{ {F^{AD}}_{IR} (t) = F_{OP}(t) (\nu_{IR}/\nu_{OP})^a}
    \label{eq:adf}
\end{equation}

\begin{equation}
    \mathrm{ {F^{DUST}}_{IR} (t) = {F^{Observed}}_{IR} (t) - {F^{AD}}_{IR} (t) }
    \label{eq:adc}
\end{equation}

where $F^{AD}_{IR}(t)$ is the accretion disk contribution to the IR flux and $F_{OP}(t)$ is the optical flux in the V-band and $\nu_{IR}$ and $\nu_{OP}$ are the frequency corresponding to the IR and optical bands, respectively. Here, $a$ is the unknown power-law index, which is assumed to be  1/3, considering the standard model \citep{1973A&A....24..337S}, and from the observational support that accretion disk power-law
continuum extends into the near-IR until about 2 $\mu$m \citep{2008Natur.454..492K}. Previous studies have used different values of $a$. For example, \citet{2019ApJ...886..150M} used $a$=0.1, \citet{lyu2019mid} used $a$=1/3, while \citet{yang2020dust} and \citet{chen2023mid} did not perform any accretion disk correction in their dust lag analysis. 

Recently, \citet{mandal2024revisiting}, using a subsample of their objects, found that the dust lag increases insignificantly (0.03 dex only) if $a$ =0 is used instead of 1/3. They found a slope of 1.12 between accretion disk-corrected and uncorrected dust lags, suggesting a possible greater accretion disk contamination in high-luminosity AGNs; however, no correlation with accretion rate was observed. Following \citet{mandal2024revisiting}, we corrected the accretion disk contribution using the above-mentioned equations \ref{eq:adf} and \ref{eq:adc}. For this, to calculate the optical flux at IR epochs, we performed DRW modeling of the optical light curves. In Figure \ref{fig:adc}, we compared the accretion disk corrected vs. non-corrected lag in W1 and W2 bands. We found that the disk contamination has an insignificant effect on the lag in our sample, with less than 1\% increase in both bands, with a slope of 1.12 between accretion disk corrected vs. uncorrected lags. Additionally, we observed that the W1 band is more contaminated by accretion disk emission than the W2 band, as expected. However, to be consistent with the recent studies \citep[e.g.,][]{mandal2024revisiting}, we have adopted the accretion disk corrected dust lags in all further analyses. We tabulated the lag results for all the objects obtained via the ICCF method in the table \ref{tab:lag-result}.

\begin{figure*}
    \centering
    \includegraphics[width=8.5cm,height=8cm]{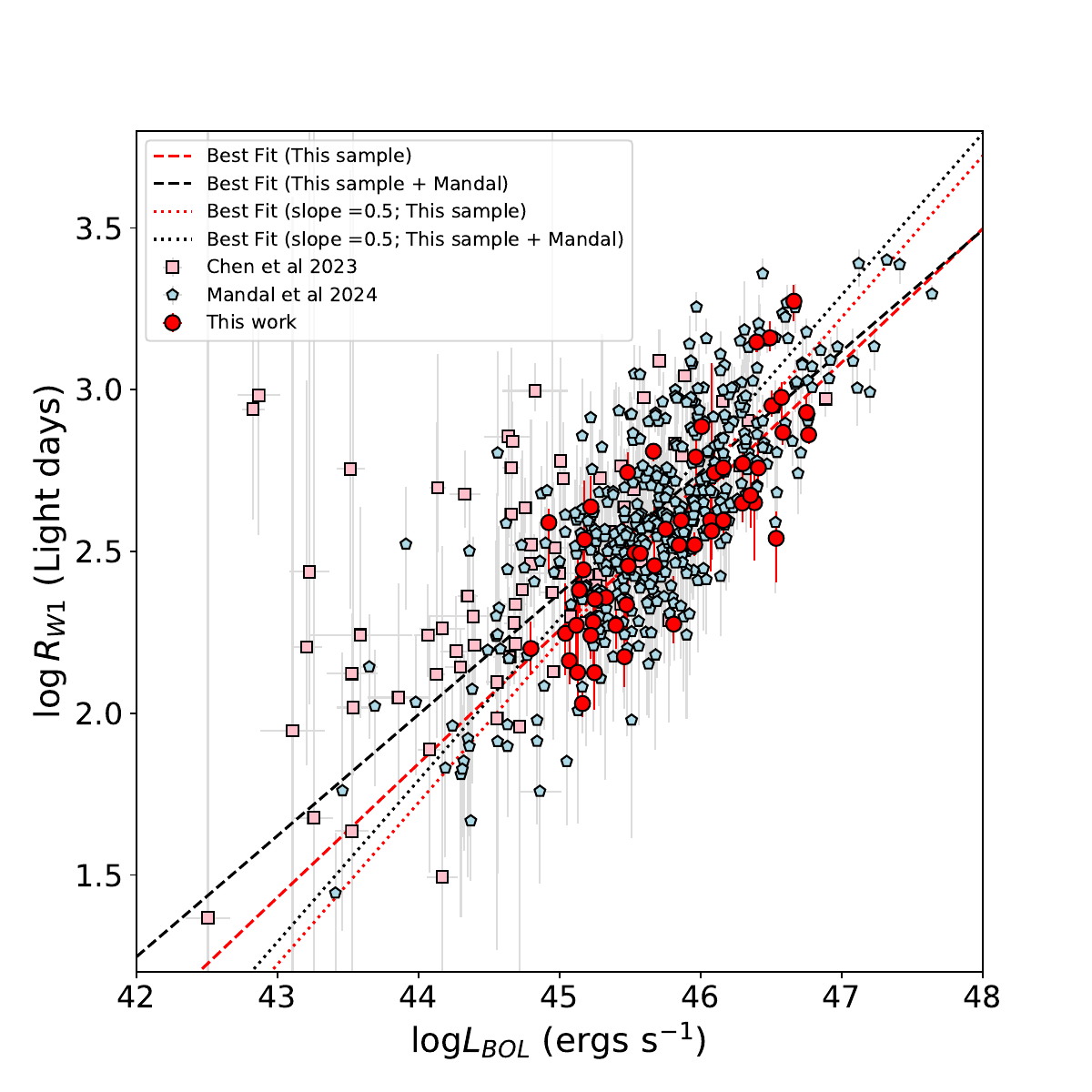}
    \includegraphics[width=8.5cm,height=8cm]{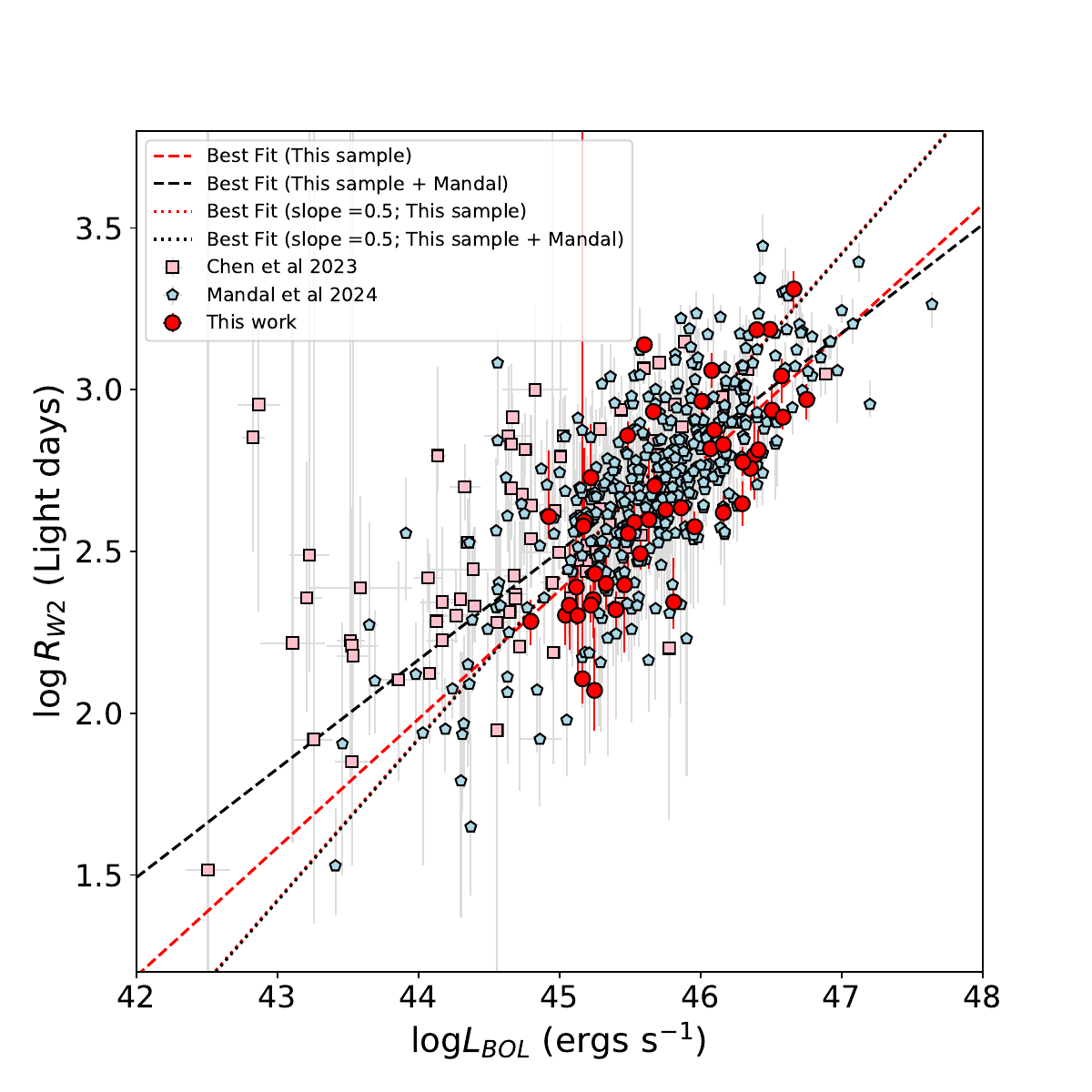}
    \caption{The relation between torus radius vs. bolometric luminosity in W1 (left) and W2 (right) bands. The torus radii from \citet{chen2023mid} and \citet{mandal2024revisiting} are also shown, noting that the values from the former were not corrected for accretion disk contamination. The best-fit relations for this sample and the one including \citet{mandal2024revisiting} sample are shown with different colors. Note that the sample from \citet{chen2023mid} is not included in the best-fit analysis since AD contamination has not been corrected in that case.}
    \label{fig:RL_fixed_slope}
\end{figure*}

\section{Results and Discussion} \label{ss:results}

\subsection{The relation between the torus size and AGN luminosities}

In this section, we examine the relation between the measured torus size and the AGN luminosities. We have utilized the redshift and accretion disk contamination corrected time lags as a measure of torus radii, and the bolometric luminosity $L_{\text{BOL}}$ taken from \citet{rakshit2020spectral}. To calculate $L_{\text{BOL}}$, the authors performed multi-component fitting of the SDSS spectra of the objects to estimate $L_{5100}$, first subtracting host-galaxy contribution if present, then modeling the AGN continuum using power-law, Fe II emission, etc. The $L_{5100}$ obtained from the best-fit power-law component is then used with the bolometric correction factor from \citet{richards2006spectral} to calculate $L_{\text{BOL}} = 9.26 \times L_{5100}$ for $z < 0.8$.

In Figure \ref{fig:RL_fixed_slope}, we plot the size of the dust torus against its bolometric luminosity. The correlation between the torus sizes and the AGN luminosities is found by performing a linear regression analysis as

\begin{equation}
    \rm log(R_{dust}/ 1 \, light\text{-}day) = \beta + \alpha\, \log(L_{BOL}/L_0)
    \label{RLequation}
\end{equation}
where $\alpha$ and $\beta$ are the slope and intercept, respectively. $L_0 = 45.80$ erg s$^{-1}$ is the normalizing constant for the luminosity, which is set approximately equal to the median of the luminosities of the final sample. To perform the linear regression analysis, we have employed the LINMIX \citep{2007ApJ...665.1489K} and Bivariate Correlated Errors and Intrinsic Scatter \citep[BCES;][]{akritas1996linear}. Both the LINMIX and BCES are used as they account for the intrinsic scatter in addition to the measurement errors and the correlations within the measurement errors. We consider two cases in our linear regression analysis: one where both the slope and intercept are free parameters and another where the slope is fixed at $\alpha = 0.5$. The latter case is motivated by the dust radiation equilibrium model, which predicts that the dust sublimation radius scales with UV luminosity as a power law with an exponent of 0.5 using the emcee code\footnote{\url{https://emcee.readthedocs.io/en/stable/tutorials/line/}}. For the fitting with variable slope we find the torus sizes follow a power-law relation with AGN luminosity, given by $R \propto L_{\mathrm{BOL}}^{0.413\pm0.047}$ for W1 and $L_{\mathrm{BOL}}^{0.397\pm0.058}$  in W2 bands for this sample and a slightly shallower slope for the combined sample with $R_{W1} \propto L_{\mathrm{BOL}}^{0.374\pm0.015}$ and $R_{W2} \propto L_{\mathrm{BOL}}^{0.336\pm0.016}$ when fitted using LINMIX. A similar trend of shallower slope in the W2 band compared to the W1 band has been found when fitted using BCES.

We find that the slope in $R_{\mathrm{dust}}- L_{\mathrm{BOL}}$ relation (equation \ref{RLequation}) is shallower than the expected slope of 0.5 from the dust sublimation model. This can be attributed to the fact that we use AGN bolometric luminosity, $L_{\text{BOL}}$, which is derived from  $L_{5100}$ for the fitting, may not be the actual representative of the ionizing luminosity. Other reasons that contribute to the shallower slopes are the anisotropic illumination of the torus and the self-shadowing effects of the slim accretion disk. Compared to the previous works \citep[e.g.,][]{chen2023mid, mandal2024revisiting}, the slope obtained from our sample is slightly steeper, but it is consistent with the recent dust continuum observations using GRAVITY interferometric measurements \citep{2023A&A...669A..14G}.

\begin{figure}
    \centering
    \includegraphics[width=1\linewidth,height=1\linewidth]{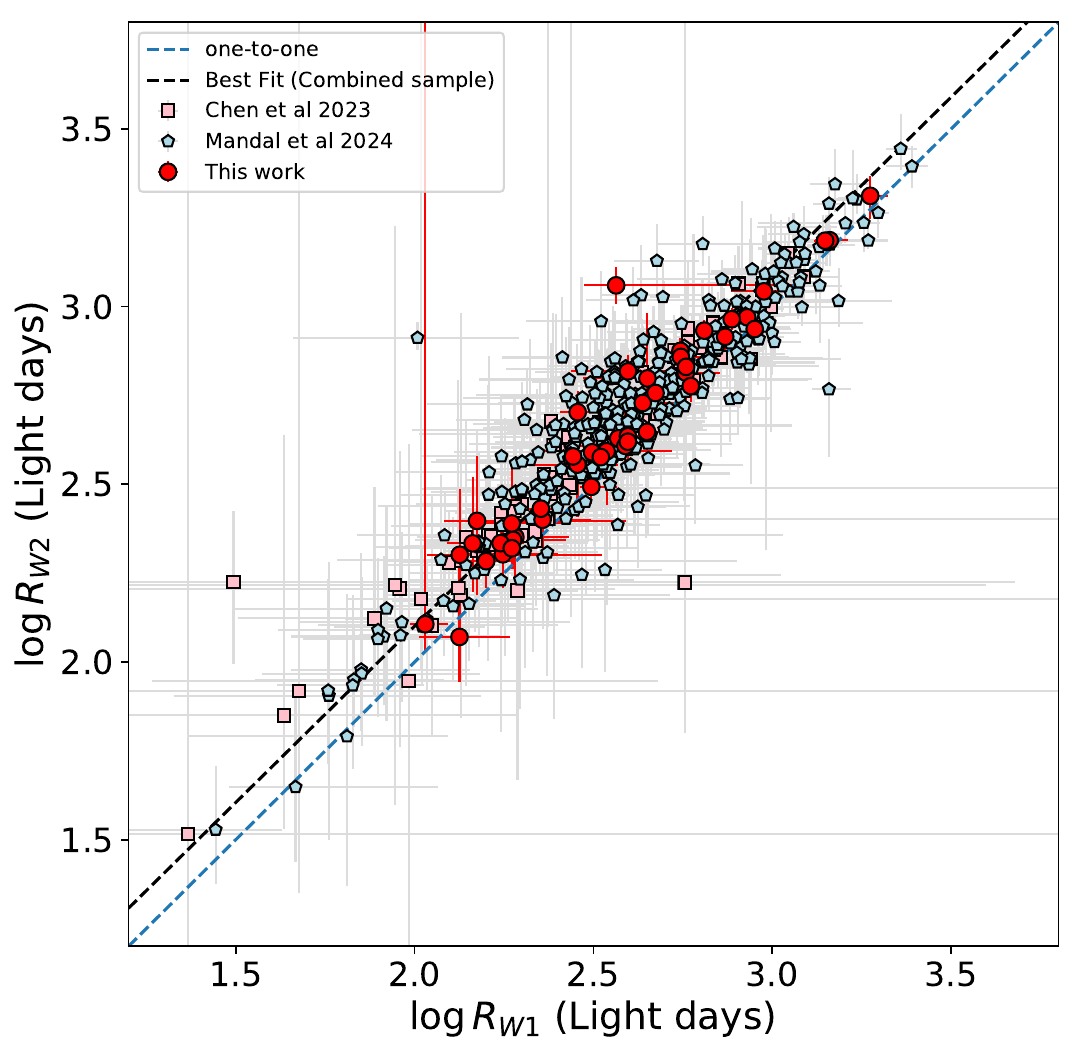}
    \caption{Comparison of the W1 and W2 lags after redshift corrections. The dashed black line represents the best-fit, while the 1:1 relation is shown by the dashed blue line.}
    \label{fig:w1_w2 analysis}
\end{figure}

From our sample, a total of 45 sources have lag measurement in both W1 and W2 bands. We plot the W2 vs W1 rest-frame lags in Figure \ref{fig:w1_w2 analysis} and find that the mean ratio of the torus sizes in the W2 and W1 bands is 1.17, which states that the torus sizes in W2 are greater than those obtained in the W1 band. We also include the objects from \citet{chen2023mid} and \citet{mandal2024revisiting} in the same figure. The ratio of W2 to W1 band lag for the combined sample is 1.21. Furthermore, we perform BCES orthogonal regression analysis and find
\begin{equation}
    \log R_{W2} = (0.99 \pm 0.12)\times \log R_{W1} + (0.12\pm 0.30)
\end{equation}
Thus, the results are consistent with the expectation that the torus sizes increase with increasing wavelength, indicating a stratified torus \citep{mandal2024revisiting}.

\begin{figure}
    \centering
    \includegraphics[width=9cm,height=5cm]{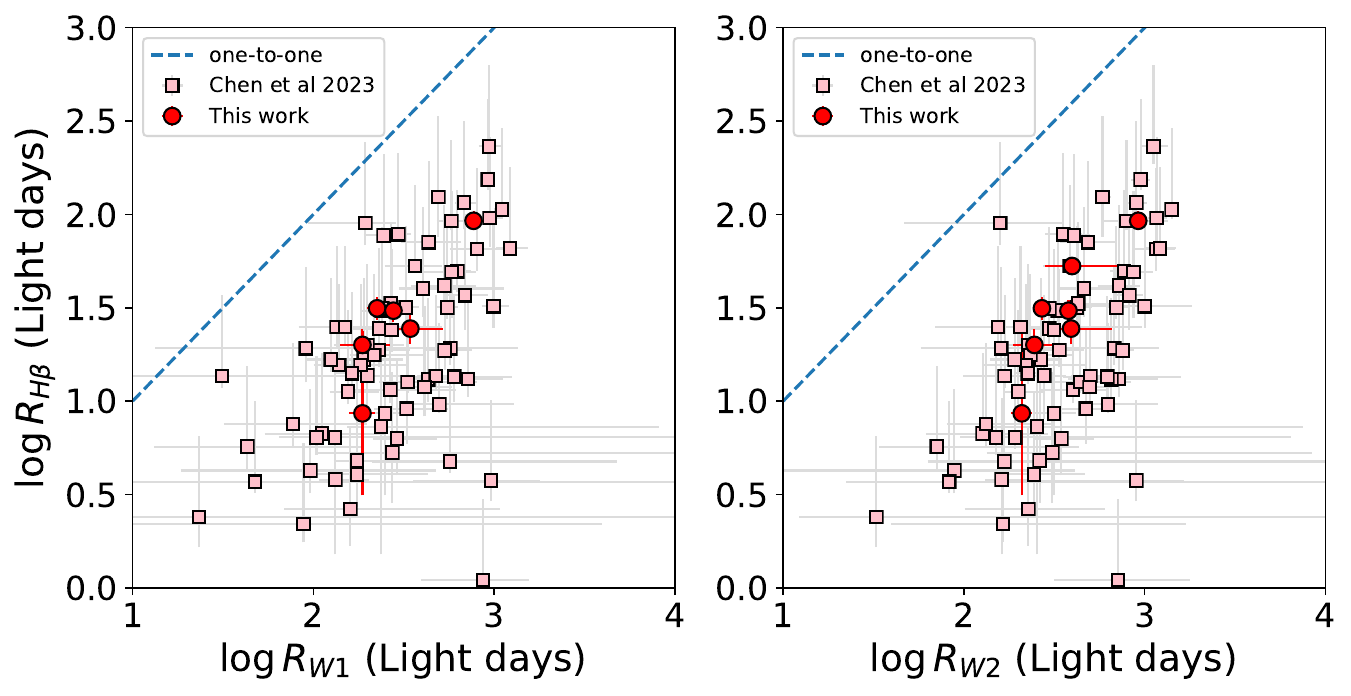}
    \caption{Comparison of BLR and dust torus sizes in W1 (left) and W2 (right) bands. Our measurements are shown as circles, while the squares represent the measurements from \citet{chen2023mid}. A one-to-one line is shown for reference.}
    \label{fig:com_blr_dust}
\end{figure}

\begin{table*}
\centering
\caption{Results of the linear regression fit to the torus radius -- luminosity relation.}
\label{tab:best-fitting}
    \begin{tabular}{ccccccccc}
        \hline
        \hline
        Sample & Band & $\alpha_{\mathrm{free}}$ & $\beta_{\mathrm{free}}$ & $\sigma_{\mathrm{free}}$ & $\alpha_{\mathrm{free}}$ & $\beta_{\mathrm{free}}$ &    $\alpha_{\mathrm{fixed}}$  & $\beta_{\mathrm{fixed}}$ \\
            &     & \multicolumn{3}{|c|}{LINMIX} & \multicolumn{2}{c|}{BCES} & \multicolumn{2}{c|}{EMCEE}  \\  
        (1) & (2) & (3) & (4) & (5) & (6) & (7) & (8) & (9) \\
        \hline
    
        This Work & W1 &  $0.413 \pm 0.047$ & $2.588 \pm 0.026$ & 0.16  & 0.399 $\pm$ 0.046 & 2.573 $\pm$ 0.025  & 0.5 & 2.624 $\pm$ 0.014  \\
                  & W2 & $0.397 \pm 0.058$ & $2.698 \pm 0.030$ & 0.19  & 0.405 $\pm$ 0.047 & 2.681 $\pm$ 0.028 & 0.5 & 2.823 $\pm$ 0.016  \\
        This Work + Mandal & W1 &  $0.374 \pm 0.015$  & $2.670 \pm 0.009$ & 0.16  &  0.383 $\pm$ 0.016 & 2.642 $\pm$ 0.009 &  0.5 & 2.693 $\pm$0.009 \\
                       & W2 & $0.336 \pm 0.016$  & $2.770 \pm 0.009 $ & 0.16  & 0.352 $\pm$ 0.017 & 2.746 $\pm$ 0.008 & 0.5 & 2.819 $\pm$ 0.012  \\
        \hline
    \end{tabular}

        \raggedright  Note: Col. (1): sample used; Col. (2): infrared bands for which lags were measured relative to the optical band; Cols. (3–5): best-fit slope, intercept, and intrinsic scatter from LINMIX regression; Cols. (6–7): best-fit slope and intercept from BCES regression; Cols. (8–9): best-fit intercept values with slope fixed to 0.5 from emcee fitting.
\end{table*}

\subsection{Comparison between BLR and Torus sizes.}

According to the unified model of AGNs \citep{1995PASP..107..803U}, the dusty torus is expected to lie beyond the BLR. Several studies have explored the scaling relation between the BLR and torus sizes. For instance, \citet{du2015supermassive} compared their results with \citet{koshida2014reverberation} and found that for low-accretion AGNs, the ratio \( R_{\mathrm{dust,K}}/R_{\mathrm{H}\beta} \sim 4 \) based on a sample of 10 sources. For high-accretion AGNs, this ratio increased to \(\sim 7\), albeit from a smaller sample of just four sources. Similarly, \citet{kokubo2020rapid} analyzed 15 Seyfert galaxies and reported that the torus typically lies at a distance roughly four times that of the BLR.

In a more comprehensive analysis, \citet{chen2023mid} derived scaling relations for 78 AGNs and reported \( R_{\mathrm{dust,K}}/R_{\mathrm{H}\beta} = 6.2 \), \( R_{\mathrm{dust,W1}}/R_{\mathrm{H}\beta} = 9.2 \), and \( R_{\mathrm{dust,W2}}/R_{\mathrm{H}\beta} = 11.2 \). These results were broadly supported by \citet{mandal2024revisiting}, who reported \( R_{\mathrm{dust,W1}}/R_{\mathrm{H}\beta} = 0.98 \pm 0.36 \) dex (corresponding to a size ratio of \(\sim 9.5\)) and \( R_{\mathrm{dust,W2}}/R_{\mathrm{H}\beta} = 1.18 \pm 0.36 \) dex (\(\sim 15\)), based on a sample of 19 AGNs with reverberation-mapped BLR sizes. The authors noted that the relatively larger scatter in W2 may be due to the smaller sample size compared to \citet{chen2023mid}.

In our study, we further investigate the relationship between BLR and torus sizes in the W1 and W2 bands using recent H\(\beta\) lag measurements from \citet{2024ApJS..275...13W}, who performed a uniform analysis of archival light curves. Cross-matching with our sample yields 6 (W1) and 7 (W2) objects with 
H$\beta$ lags. We plot the BLR size against the dust torus size in W1 and W2 bands in Figure \ref{fig:com_blr_dust}. We find a relatively shallow scaling relation: 
$R_{\mathrm{BLR}} : R_{\mathrm{W1}} : R_{\mathrm{W2}} = 1.0 : 9^{+6}_{-1} : 12^{+4}_{-4}$. To expand the sample, we also cross-matched the \citet{chen2023mid} sample with the H$\beta$ lags from \citet{2024ApJS..275...13W}, identifying 72 sources with both measurements. For this larger sample, we find $R_{\mathrm{BLR}} : R_{\mathrm{W1}} : R_{\mathrm{W2}} = 1.0 : 12^{+23}_{-6} : 17^{+30}_{-9}$, which shows larger ratios than those originally reported by \citet{chen2023mid}. This discrepancy may arise because multiple lag measurements exist for the same object with significant variation. We adopted the average H\(\beta\) lag per object. Moreover, the updated lags from \citet{2024ApJS..275...13W} are derived using a uniform and systematic methodology. For example, for PG1552+085, \citet{2021ApJS..253...20H} reported a BLR lag of 25 days, while using the same data, \citet{2024ApJS..275...13W} reported a revised value of just 8.6 days. Despite these caveats, our findings reinforce the conclusion that the dust torus sizes in W1 and W2-bands, are significantly larger than the BLR, consistent with the AGN unification paradigm.

\begin{figure*}
    \centering
    \includegraphics[width=8cm,height=8cm]{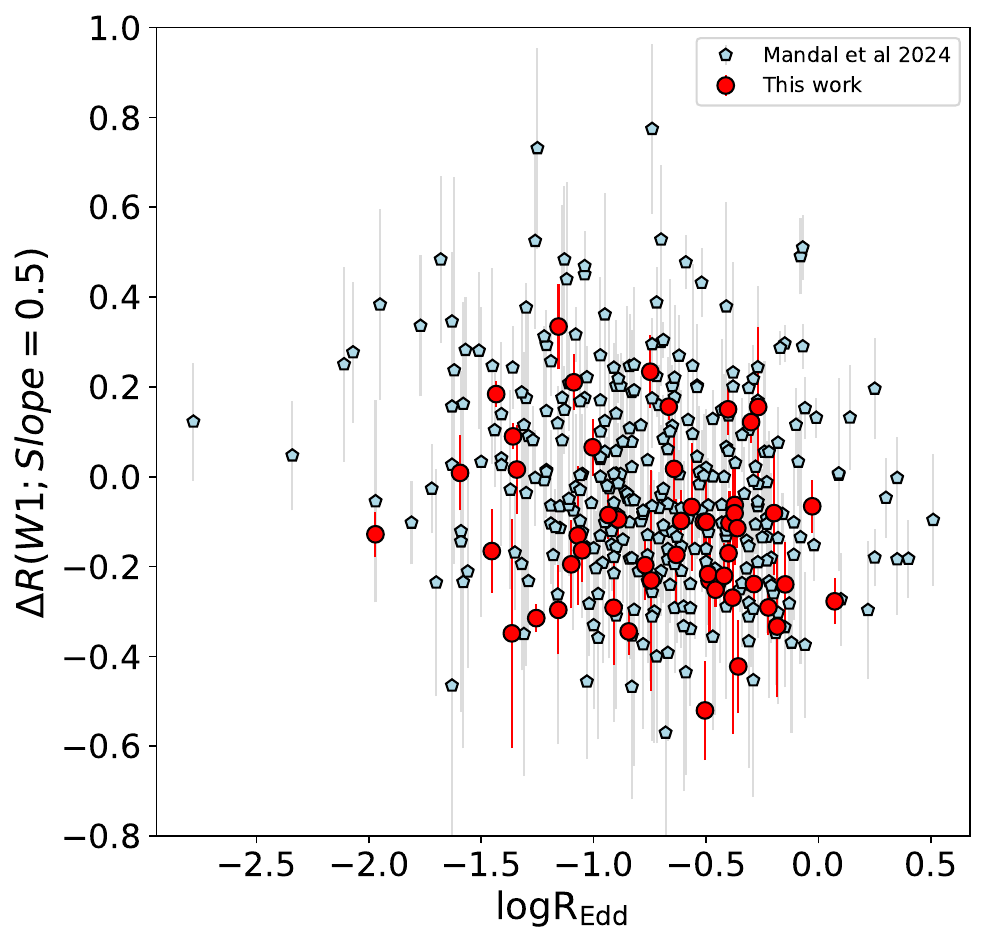}
    \includegraphics[width=8cm,height=8cm]{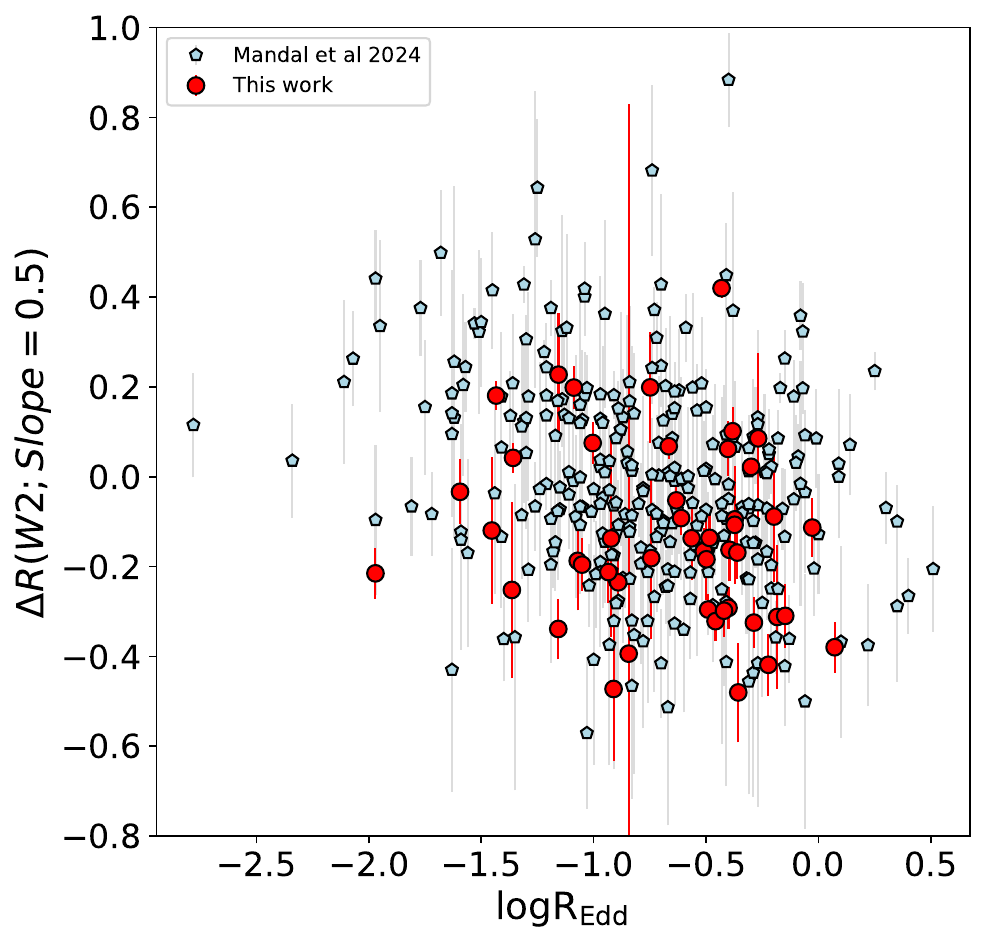}
    \caption{The correlation of $\Delta R (\log R_{\mathrm{obs}} - \log R_{\mathrm{best-fit, slope=0.5}})$ with Eddington ratio for W1 band (left) and W2 band (right). The measurements from this work are shown as red points, while the blue points are from \citet{mandal2024revisiting}.}
    \label{fig:lag_redd}
\end{figure*}

\begin{figure*}
    \centering
    \includegraphics[width=18cm,height=4.5cm]{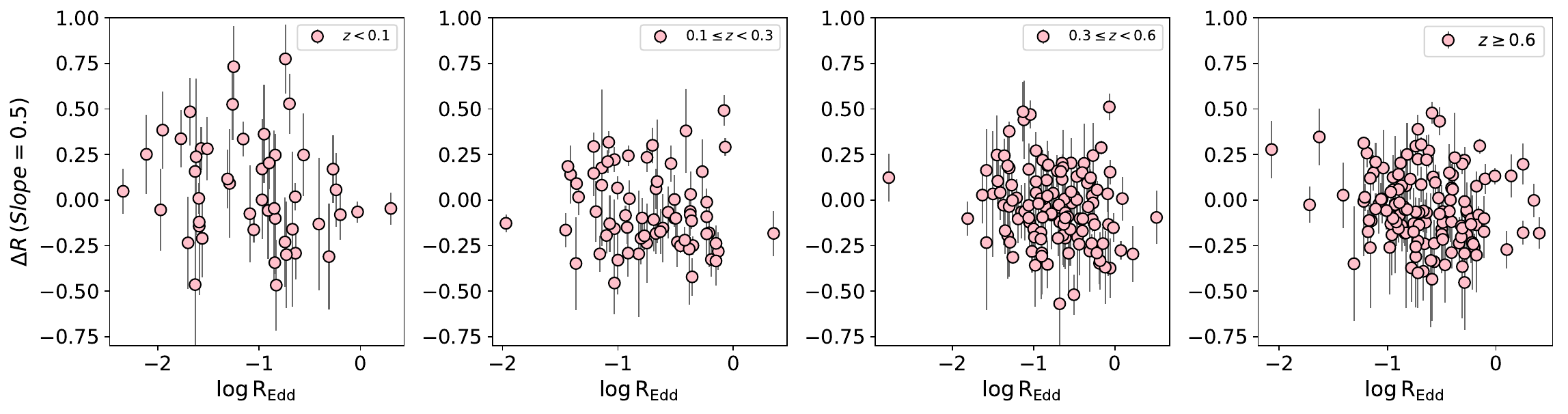}
    \includegraphics[width=18cm,height=4.5cm]{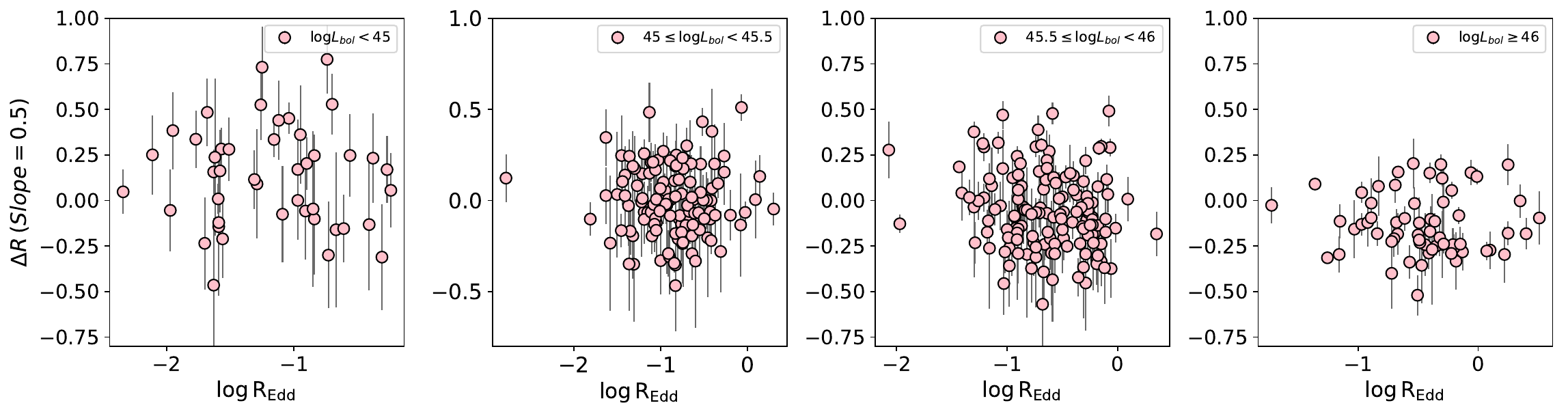}
    \caption{The correlation of $\Delta R$ with Eddington ratio for W1 band in different bins of redshift and bolometric luminosity for the combined sample (this work and \citet{mandal2024revisiting}) in the top and bottom panel, respectively.}
    \label{fig:lag_redd_bins}
\end{figure*}

\begin{table}
\centering
\movetableright= -15mm
\caption{The correlation of $\Delta R$ with Eddington ratio for different redshift and bolometric luminosity bins}
\label{tab:corr}
\resizebox{9cm}{!}{%
    \begin{tabular}{cllll}
        \hline \hline
        Parameter bins & r\_s  & p-value  & $\tau$ & $p_{\tau}$ \\
                &     \multicolumn{2}{|c|}{Spearman}       &  \multicolumn{2}{|c|}{Kendall tau} \\
        (1)     &     (2)       & (3)      & (4)   & (5)  \\\hline
        $z<0.1$     & -0.23 (-0.23) & 0.11 (0.11) & -0.15 (-0.13) & 0.13 (0.19)\\
        $0.1<z<0.3$ & -0.18 (-0.17) & 0.14 (0.16) & -0.13 (-0.18) & 0.13 (9e-7)\\
        $0.3<z<0.6$ & -0.19 (-0.31) & 0.03 (0.001) & -0.14 (-0.16) & 0.03 (0.11) \\
        $z>0.6$     & -0.19 (-0.19) & 0.02 (0.03)  & -0.13 (-0.12) & 0.03 (0.15) \\   \hline
        log $L_{BOL}<45$     & -0.08 (-0.03) & 0.56 (0.85) & -0.06 (-0.21) & 0.57 (9e-4)\\
        $45<\text{log} L_{BOL}<45.5$ & -0.04 (-0.23) & 0.60 (0.01) & -0.03 (-0.14) & 0.61 (0.02) \\
        $45.5<\text{log} L_{BOL}<46$ & -0.15 (-0.12 )& 0.07 (0.15) & -0.11 (-0.09) & 0.05 (0.14) \\
        log $L_{BOL}>46$      & -0.06 (-0.07) & 0.62 (0.57) & -0.06 (-0.06) & 0.52 (0.53) \\
        \hline
 \end{tabular}}
 \raggedright  Note: Col. (1): redshift and luminosity bins, Cols. (2-3): Spearman rank correlation coefficient and probability value of no correlation for W1, Cols. (4-5): Kendall tau correlation coefficient and p-value of no correlation (the values in the parentheses are for the W2 band).
\end{table}

\subsection{Dependency of Torus sizes on accretion rates}

Studies of super-Eddington accreting massive black holes (SEAMBHs; \citealt{du2014supermassive}, \citealt{du2015supermassive}, \citealt{2018ApJ...856....6D}, \citealt{2014ApJ...797...65W}, \citealt{2021ApJS..253...20H}) have consistently shown that H$\beta$ time lags measured with respect to the continuum at 5100\,\AA\ are systematically shorter in high-accretion AGNs than in their sub-Eddington counterparts. This suppression in lag can be attributed to several factors: (1) the self-shadowing effect of slim accretion disks, which are optically and geometrically thick and exhibit a funnel-like structure \citep[e.g.,][]{2014ApJ...793..108W}; (2) the impact of black hole spin on disk structure and radiation \citep[e.g.,][]{2019ApJ...870...84C}; and (3) variations in the ionizing continuum and UV/optical spectral energy distribution \citep{2020ApJ...899...73F}.

To investigate the scatter observed in the $R_{\text{dust}}$--$L_{\text{BOL}}$ relation, we define the deviation from the expected radius--luminosity relation with a fixed slope of 0.5 as:
\begin{equation}
    \Delta R = \log R_{\rm obs} - \log R_{\rm best-fit,~slope=0.5}.
\end{equation}

We adopt Eddington ratio values from the catalog of \citet{rakshit2020spectral}, where the Eddington luminosity is computed as \(L_{\rm Edd} = 1.26 \times 10^{38}~(M_{\rm BH}/M_\odot)\,\rm erg\,s^{-1}\), and the Eddington ratio as \(L_{\rm BOL}/L_{\rm Edd}\). In Figure~\ref{fig:lag_redd}, we plot the deviations \(\Delta R_{W1}\) and \(\Delta R_{W2}\) in the W1 and W2 bands, respectively, against the Eddington ratio. It is important to note that these Eddington ratios are derived using single-epoch black hole mass estimates, which are known to carry substantial uncertainties that may bias the results \citep[see discussion in][]{mandal2024revisiting}.

To assess the robustness of these correlations, we perform a Spearman rank-order correlation and Kendall's tau analysis. For our sample, we find in the W1 band a correlation coefficient of \(r_s = -0.17\) (p-value = 0.23), and for the combined sample (this work + \citealt{mandal2024revisiting}), \(r_s = -0.22\) (p-value = $10^{-5}$).  For the W2 band, the correlation coefficients are \(r_s = -0.19\) (p-value = 0.189) and \(r_s = -0.26\) (p-value = $9 \times 10^{-7}$), respectively. Similarly, Kendall's tau analysis provides $\tau=-0.12$ (${p_{\tau}}=0.19$) for W1 and $\tau=-0.13$ (${p_{\tau}}=0.19$) for W2 bands for our sample, and $\tau=-0.15$ (${p_{\tau}}=1e-5$) for W1 and $\tau=-0.19$ (${p_{\tau}}=9e-7$) for W2 bands for the combined sample. These modest negative trends may reflect the underlying coupling between the BLR and the dust-reverberation radius, particularly in light of the known correlation between \(R_{\rm H\beta}\) and \(R_{\rm IR}\). Since the deviations scale as \(\Delta R_{\rm IR} \propto R_{\rm dust,IR}/L^{0.5}\) \citep{mandal2024revisiting, alvarez2020sloan}, the observed trend with Eddington ratio may arise as an indirect effect of this dependence.

To further explore these dependencies, we divide the sample into different bins of bolometric luminosity and redshift (Figure \ref{fig:lag_redd_bins}). The correlation coefficients ($|{r_s}|<0.2$) indicate weak trends in all redshift bins (see Table \ref{tab:corr}); the associated $p$-values differ because of the varying sample sizes. Moreover, when the sample is divided by bolometric luminosity, the correlation nearly vanishes. Our results suggest that torus size deviations, like those of the BLR, mildly anti-correlate with Eddington ratio across redshift. This trend weakens with increasing bolometric luminosity, suggesting that accretion rate, rather than luminosity itself, plays a more significant role in shaping the inner structures of AGNs.

\section{Summary}\label{ss:sum}

We present the dust reverberation mapping results of a sample of objects selected from \citet{rakshit2020spectral} utilizing the data from survey telescopes, such as ASAS-SN, CRTS, ZTF, and WISE/NEOWISE. We have employed ICCF and MICA to estimate the lag. Our initial sample, which consists of 146 objects, is reduced to 51 objects for the W1 band and 47 objects for the W2 band after performing the quality assessment of the lag obtained from the light curves of the objects by employing PyIICCF. Finally, the lags obtained are corrected for accretion disk contamination and redshift effect to account for the redshift dependence of the observed lags and to ensure that the lags obtained are in the rest frame. With the final set of objects, we have explored the R$_{dust, IR}$-L$_{BOL}$ relation, and our major findings are summarized below.

\begin{itemize}

\item We find that the torus radii scales with luminosity as R$_{dust, W1}$ $\propto $L$_{BOL}^{0.413 \pm 0.047}$ and R$_{dust, W2}$ $\propto $L$_{BOL}^{0.397 \pm 0.058}$ with intrinsic scatter of 0.16 and 0.19, respectively, when the slope is treated as a free parameter. However inclusion of literature samples from \citet{mandal2024revisiting}  particularly affects the W2 band and the slopes for the combined sample become shallower, resulting R$_{dust, W1}$ $\propto $L$_{BOL}^{0.374 \pm 0.015}$ and R$_{dust, W2}$ $\propto $L$_{BOL}^{0.336 \pm 0.016}$ with intrinsic scatter of 0.16. These slopes are significantly shallower than the R$_{dust, IR}$ $\propto $L$_{BOL}^{0.5}$ relation predicted by the dust sublimation equilibrium model. However, they are consistent with the torus radius -- luminosity relations observed from IR-interferometry, which similarly indicate a shallower dependence compared to the theoretical dust sublimation scenario.

\item When the slope is fixed at 0.5, we obtain intercepts of 2.624 $\pm$ 0.014 for the W1 band and 2.823 $\pm$ 0.016 for the W2 band. The dust torus sizes in the W1 and W2 bands exhibit a strong correlation, with a time-lag ratio of $\tau_{W2}/\tau_{W1} = 1.21 \pm 0.06$, following the relation $R_{W2}=(0.99\pm0.12)\times \log R_{W1} + (0.12\pm0.30)$. Furthermore, the characteristic sizes of various AGN components are found to follow the ratio R$_{BLR}$: R$_{W1}$: R$_{W2}$ = 1: $9^{+6}_{-1}$: $12^{+4}_{-4}$, indicating a stratified torus structure that lies beyond the BLR.

\item We find a modest negative correlation between the deviation of the dust torus radius from the best-fit relation with a fixed slope of 0.5 and the accretion rate, as characterized by the Eddington ratio. This correlation persists across different redshift bins but disappears when the data are divided into luminosity bins. These results indicate a possible influence of the accretion rate on the dust torus radius, potentially due to self-shadowing effects of the accretion disk, which may contribute to a flattening of the dust radius -- luminosity relation.
\end{itemize}

\begin{acknowledgments}
We dedicate this paper to the memory of Ashutosh Tomar (1998–2025), who led this work with clarity, dedication, and deep scientific insight. Ashutosh passed away before the submission of this manuscript. We deeply miss his presence as a colleague and friend.

A.K.M. acknowledges the support from the European Research Council (ERC) under the European Union’s Horizon 2020 research and innovation program (grant No. 951549).
This publication uses data products from the WISE, a joint project between the University of California, Los Angeles, and the Jet Propulsion Laboratory/California Institute of Technology, and is funded by the National Aeronautics and Space Administration. The publication also leverages data products from NEOWISE, a Jet Propulsion Laboratory/California Institute of Technology project, funded by the Planetary Science Division of the National Aeronautics and Space Administration. The CRTS survey is conducted by the U.S. National Science Foundation under grants AST-0909182 and AST-1313422. The ASAS-SN is supported by the Gordon and Betty Moore Foundation through grant GBMF5490 to the Ohio State University and NSF grant AST-1515927. The Development of ASAS-SN was supported by the NSF grant AST- 0908816, the Mt. Cuba Astronomical Foundation, the Center for Cosmology and AstroParticle Physics at the Ohio State University, the Chinese Academy of Science South America Center for Astronomy (CASSACA), the Villum Foundation, and George Skestos. ZTF is funded by the National Science Foundation through grant No. AST-1440341 and a partnership with Caltech, IPAC, the Weizmann Institute for Science, the Oskar Klein Center at Stockholm University, the University of Maryland, the University of Washington, Deutsches Elektronen-Synchrotron and Humboldt University, Los Alamos National Laboratories, the TANGO Consortium of Taiwan, the University of Wisconsin at Milwaukee, and Lawrence Berkeley National Laboratories. COO, IPAC, and UW manage operations. This paper also used data observed with the Samuel Oschin Telescope at the Palomar Observatory as part of the Palomar Transient Factory project, a scientific collaboration between the California Institute of Technology, Columbia University, Las Cumbres Observatory, the Lawrence Berkeley National Laboratory, the National Energy Research Scientific Computing Center, the University of Oxford, and the Weizmann Institute of Science.
\end{acknowledgments}


\facilities{CRTS, ZTF, ASAS-SN, WISE}
\software{{\tt PyCALI} \citep{li2014bayesian}, MICA\citep{li2016non}, {\tt PyI$^2$CCF}, {\tt PyCCF} \citep{2018ascl.soft05032S}}

\bibliography{biblio.bib}
\bibliographystyle{aasjournalv7}






\label{lastpage}
\end{document}